\documentclass[%
 reprint,
nofootinbib,
 amsmath,amssymb,
 aps,
]{revtex4-1}
\usepackage{slashbox}
\usepackage{array}
\usepackage{nameref}
\usepackage{xcolor}
\usepackage{graphicx}
\usepackage{dcolumn}
\usepackage{bm}

\begin{document}

\title{Heavy Particle Jet Identification with Zest}

\author{Ankita Budhraja}
\email{ankitab@iiserb.ac.in}
\author{Ambar Jain}%
 \email{ambarj@iiserb.ac.in}
\affiliation{%
 Indian Institute of Science Education and Research, Bhopal MP 462066, India
}

\date{\today}

\begin{abstract}

We introduce a new jet observable {\em zest} defined on exclusively constructed jets and study its potential to discriminate jets originated from Standard Model heavy particles like $W,~Z$ bosons and top quark from gluon initiated jets. Zest exhibits properties such as boost invariance, stability against global color exchange among partons, and inclusion or exclusion of a few soft particles in the jet. We also observe that for gluon jets, zest distribution is mostly insensitive to the jet mass. These properties make zest a suitable candidate for vetoing gluon jets at the colliders. Zest when used in conjunction with other substructure observables that are uncorrelated to it can further improve gluon jet veto. We generalize zest and show that in one limit it is synonymous to particle multiplicity and in the other limit, it projects only the leading particle. Optimization on the parameter of generalized zest further improves the discrimination ability of the observable. We find that for the top quark-initiated jets, the discrimination provided by generalized zest is in close comparison with a class of machine learning-based top taggers. We propose that studying other non-linear infrared and collinear unsafe observables may help in unveiling the hidden physics of machine learning-based observables.

\end{abstract}

\maketitle

\section{Introduction}
\label{Intro}

New physics searches are often accomplished by looking for excess in cross-section pertaining to specific decay channels of new physics particles. Constraining the measured cross-section to specific decay channels can often reduce the background significantly enough to identify the excess with high confidence level. Thus, it is usually cleaner to search for new physics particles in the leptonic or photonic channels as they offer very small background at the Large Hadron Collider (LHC). However, the cleaner channels often have extremely small cross-section that it makes the searches for new physics difficult or requires a large amount of data to get a significant sample of the signal events. Additionally, leptonic searches can be difficult if the physics or decay channels of the new particle are not known. Further difficulties may arise if there is a missing transverse energy in a leptonic channel. Similar difficulties also arise for the Standard Model (SM) heavy particles, namely, $W,~Z$ bosons, Higgs boson 
and top quarks, except that the physics of these particles is well known. These heavy particles dominantly decay to light quarks and gluons (collectively partons) which then shower and fragment into jets of hadrons. Since jets are invariably produced at the LHC due to hard scattering of the constituent partons, heavy SM particles are often faked by light quark/gluon radiations and radiations coming from other sources, such as initial state radiation, underlying event\cite{Barnafoldi:2011ad, Cacciari:2009dp} and pile-up \cite{Krohn:2013lba}.
At present collider energies, these heavy SM particles are often produced with a large Lorentz boost factor and their hadronic decay products are realised as a single collimated 'fat jet'~\cite{Adams:2015hiv, Altheimer:2013yza, Abdesselam:2010pt, Altheimer:2012mn}. In such cases, the mass of the heavy particle should be reflected in the mass of the fat jet. One may expect to disentangle the signal from the large background of light parton jets, using the jet mass cut\footnote{In this work, we mainly focus on gluon jets as the background since their cross-section is much larger.}, but as it happens to be the case, gluon jet cross-section has a significant long tail even after accounting for underlying event \cite{Barnafoldi:2011ad, Cacciari:2009dp} and pile-up \cite{Krohn:2013lba}, which can be dealt with effectively using the jet grooming techniques, such as soft drop~\cite{Larkoski:2014wba, Marzani:2017kqd, Dreyer:2018tjj}, pruning \cite{Ellis:2009su, Ellis:2009me}, trimming \cite{Krohn:2009th} and mass drop/filtering \cite{Dasgupta:2013ihk, Butterworth:2008iy, Marzani:2017mva}.
Further discussion on underlying event and pile-up is beyond the scope of this work. As far as this paper is concerned, we will assume that we start with a well-groomed jet. Since a simple mass cut is not effective in identifying the particle originating the jet inspite of grooming, thus a great deal of effort has been made in jet identification by constructing observables that can reject the parton jets and reduce the background~\cite{Altheimer:2013yza, Abdesselam:2010pt, Altheimer:2012mn, Adams:2015hiv, Larkoski:2017jix, Marzani:2019hun}. The emphasis of such observables is to improve the signal rate to mistag rate ratio. This will also be the spirit of our work. The SM heavy particles provide a perfect playground for improving on such jet identification strategies as their decay channels are well understood. The purpose of our work is to improve upon these strategies by studying a new non-linear jet observable {\em zest} dependent only on the transverse momentum distribution of particles in the jet, similar to transverse-zeal introduced in the context of jet quenching studies \cite{Gavai:2015pka}.

The remainder of the paper is organized as follows. In Sec.~\ref{Zest}, we define zest and discuss its properties. In Sec.~\ref{sec:gluzest}, we outline the features exhibited by zest distribution of gluon-initiated jets that make it suitable for vetoing the gluon jets. In Sec.~\ref{Simulation}, we discuss the details of our simulations and provide the principal results of our analysis in Sec.~\ref{zestFilter}. In Sec.~\ref{bibFilter}, we introduce another substructure observable {\it bib} which also depends only on the transverse momentum but linearly and is largely uncorrelated to zest. We contrast the two observables and perform a bi-variate analysis to improve the discrimination ability in Sec.~\ref{bivariate}. In Sec.~\ref{pzest}, we  generalize zest by introducing a real parameter $p$ that can be tuned to modify the contribution of the soft particles in the jet. The parameter $p$ can be optimized to provide improved discrimination for the heavy SM particle jets from the gluon background, as discussed in Sec.~\ref{Z} and Sec.~\ref{top}. We show that for the top quark-initiated jets, the discrimination provided by generalized zest is in close comparison to recent machine learning (ML) approaches. We propose that studying such infrared and collinear (IRC) unsafe observables may help to uncover the hidden physics behind ML approaches. Since, zest is collinear unsafe and is computable for hadronic final states only, we study its dependence on different hadronization models in Sec.~\ref{hadModel}. We conclude in Sec.~\ref{Conclusion}.

\section{Zest of a Jet}
\label{Zest}

\begin{figure}
\centering
\includegraphics[scale=0.65]{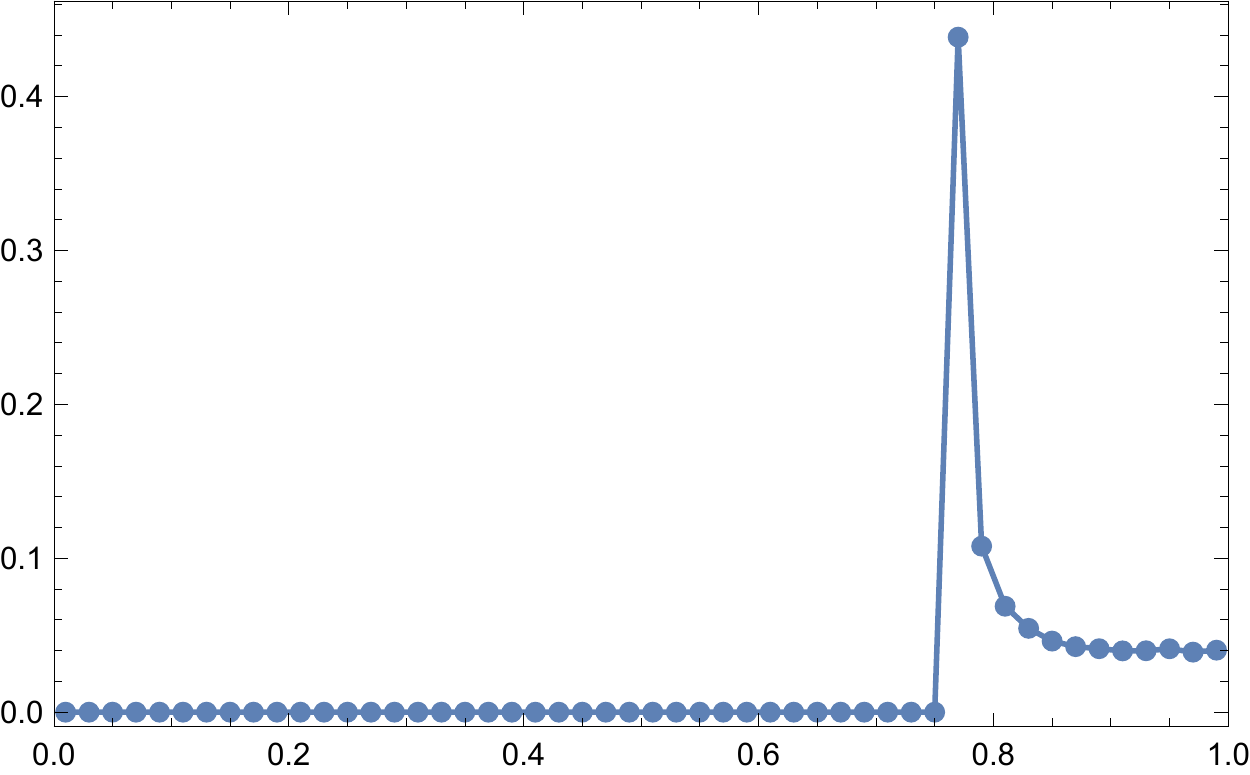}
\caption{Zest distribution for two leading particles in a jet with a fixed $P_{T}$. We see a dominant two-particle peak at a zest value of 0.76.}
\label{fig:2ptzest}
\end{figure}

Given a well-groomed jet composed of hadrons, reconstructed through a suitable jet algorithm like the anti-k$_t$ clustering algorithm~\cite{Cacciari:2008gp}, {\em zest} of the jet is defined as

\begin{eqnarray}
\label{eq:zest}
\zeta = \frac{-1}{\log \big (\sum_{i \in  \rm{Jet}} e^{-P_T/\vert  {\bf p}_{T i} \vert}\big )} \, ,
\end{eqnarray}

where $ P_{T} = \sum_{i \in \rm{Jet}} \vert \textbf{p}_{T i} \vert $ and $\textbf{p}_{T i}$ is the transverse momentum of the $i^{th}$ particle in the jet with respect to the jet axis.  
Note that zest is composed purely out of the transverse momenta of the final state particles. For the extreme case of a jet composed of a single energetic particle, it is straightforward to see that $\zeta$ reduces to $1$. Similarly, for two leading particles in the jet carrying equal fractions of energy, i.e. $P_{T} = 2\, p_{T}$, we get
\begin{equation}
\label{eq:zest2pt}
\zeta = \frac{-1}{\log(2)-2} \approx 0.765\, ,
\end{equation}  
This value remains roughly the same even if the two particles do not carry equal fractions of energy, i.e. for a generic break up of $P_{T} = x\, p_{T} + (1-x)\, p_{T}$ with $0<x<1$.
This is confirmed also by looking at FIG.~\ref{fig:2ptzest} where we have plotted the normalized distribution of $10^5$ randomly generated pairs of particles with total $P_{T}=1$. From the plot, we see that the two particle distribution has a sharp peak at about $\sim 0.77$. A similar analysis done for only three particles in a jet gives a slightly broader multiplicity peak at $\zeta \sim 0.53$. Therefore, we see that the zest distribution captures multiplicity peaks whenever there are a few energetic particles in the jet. These multiplicity peaks shift towards smaller and smaller value as the number of particles contributing to the jet increases. In the special case of a jet containing $n$ particles each carrying equal $\vert  {\bf p}_{T i} \vert$, $\zeta$ reduces to $\frac{1}{n - \log n}$. Thus we expect zest to be strongly correlated to {\em multiplicity of a jet}. Later we will generalize zest such that multiplicity will emerge as a limiting case.

Zest has certain interesting properties which makes it a useful observable to study jet substructure : (1) It is invariant under boosts made along the jet axis as it is composed entirely of transverse momentum components of the jet constituents, (2) It is sensitive to transverse momentum distribution of energetic particles while it is largely insensitive to the soft particles in the jet since their contribution to the observable is exponentially suppressed, and (3) Zest is also mostly insensitive to the global color flow of partons. In other words, the zest distribution of the gluon-initiated jets (or any other colored particle) is stable against the change of color flow direction of the colored particles forming the jet. A colored parton that evolves into a color neutral hadron jet recoils against another colored parton. The color neutral final state is obtained due to the exchange of several soft partons. Changing the direction of the recoiling colored partner effects only the exchange of a few soft particles between the partners. Since zest is stable against a few changes in the soft sector, it also becomes stable to the color recoil.

While zest is a non-linear and collinear unsafe jet observable due to which it cannot be calculated in perturbation theory, it can be computed using Monte Carlo based event generators~\cite{Sjostrand:2007gs, Bellm:2015jjp}. Although collinear unsafe, zest may provide a new perspective into the jet substructure that may not be accessible through IRC safe jet observables. In particular, the jet observables based on machine learning techniques, while not calculable through perturbation theory, provide the best discrimination ability~\cite{Kasieczka:2019dbj,Amram:2020ykb,Bradshaw:2019ipy,Dolen:2016kst,Chen:2019uar,Moreno:2019bmu,CMS:2019gpd}. The physical features extracted by such observables that enable the discrimination remains unknown. We anticipate that studying zest and other IRC unsafe observables may shed some light on the black box offered by neural-network-based observables.  It goes without saying that zest is hadronization model dependent, however we expect that its ability to discriminate between the originating particles will be independent of hadronization models used in a Monte Carlo simulator, as is demonstrated later in Sec.~\ref{hadModel}.

\section{Zest distribution of gluon-initiated jets}
\label{sec:gluzest}

As gluon jets, on an average, consist of a larger number of particles, the zest distribution of gluon-initiated jets is expected to peak at smaller values of the observable.  
\begin{figure}[t]
 \centering
  \includegraphics[width=0.45\textwidth]{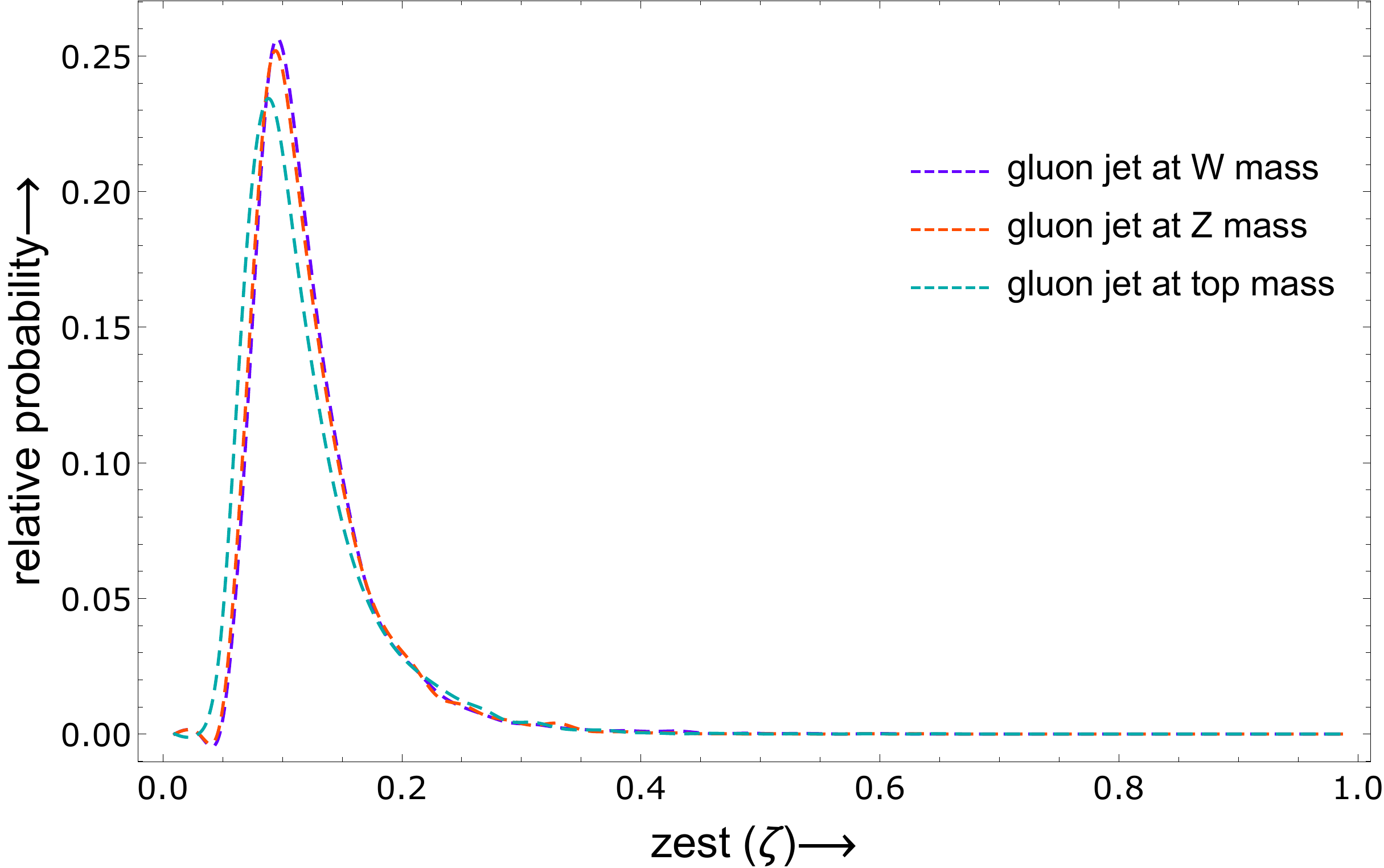}
 \caption{Zest distribution of gluon-initiated jets with a jet mass around $W$ boson mass, $Z$ boson mass and top quark mass.}
 \label{zplot}
 \end{figure}
We validate this in FIG.~\ref{zplot} by plotting zest distribution for jets initiated by offshell gluons of masses around $W,~Z$ boson and top quark masses. 
We see that the zest distribution for gluon jets peaks at small $\zeta \sim$ 0.1, as expected. Interestingly, we also see that these zest distributions are nearly independent of the jet mass. To investigate this, we have shown the ratio of the mean of $n$ maximum $p_{T i }$ to $P_T$ for $n=1,~2,~5$ and $10$ for various jet masses in TABLE~\ref{tab:veto}. We find that these ratios are nearly independent of jet mass, which explains why the zest distribution for gluons is insensitive to the jet mass.

\begin{table*}
\begin{tabular}{|m{2cm}||*{7}{c|}}\hline
\backslashbox[1.5cm]{$m_{J}$}{$n$}
&\makebox[5em]{1}&\makebox[3em]{2}&\makebox[3em]{5}
&\makebox[3em]{10}\\\hline\hline
\center gluons around W mass &0.121 $\pm$ 0.049 &0.105 $\pm$ 0.035 &0.081 $\pm$ 0.018 &0.0617 $\pm$ 0.0091\\\hline
\center gluons around Z mass &0.120 $\pm$ 0.048 &0.104 $\pm$ 0.035 &0.079 $\pm$ 0.018 &0.0603 $\pm$ 0.0094\\\hline
\center gluons around top mass &0.116 $\pm$ 0.051 &0.099 $\pm$ 0.037 &0.074 $\pm$ 0.019 &0.0552 $\pm$ 0.0102\\\hline
\end{tabular}
\\~\\
\caption{The entries in the table give the ratio of mean of $n$ maximum $\vert p_{T i}\vert$ to $P_{T}$ of the jet, that is $\frac{\langle max_{n}\vert p_{T i}\vert \rangle}{P_{T}}$. The first set of numbers specify the mean value and the second is the standard deviation calculated over a set of 10000 jets with the specified jet mass.}
\label{tab:veto}
\end{table*}

\begin{figure}[t]
 \centering
  \includegraphics[width=0.44\textwidth]{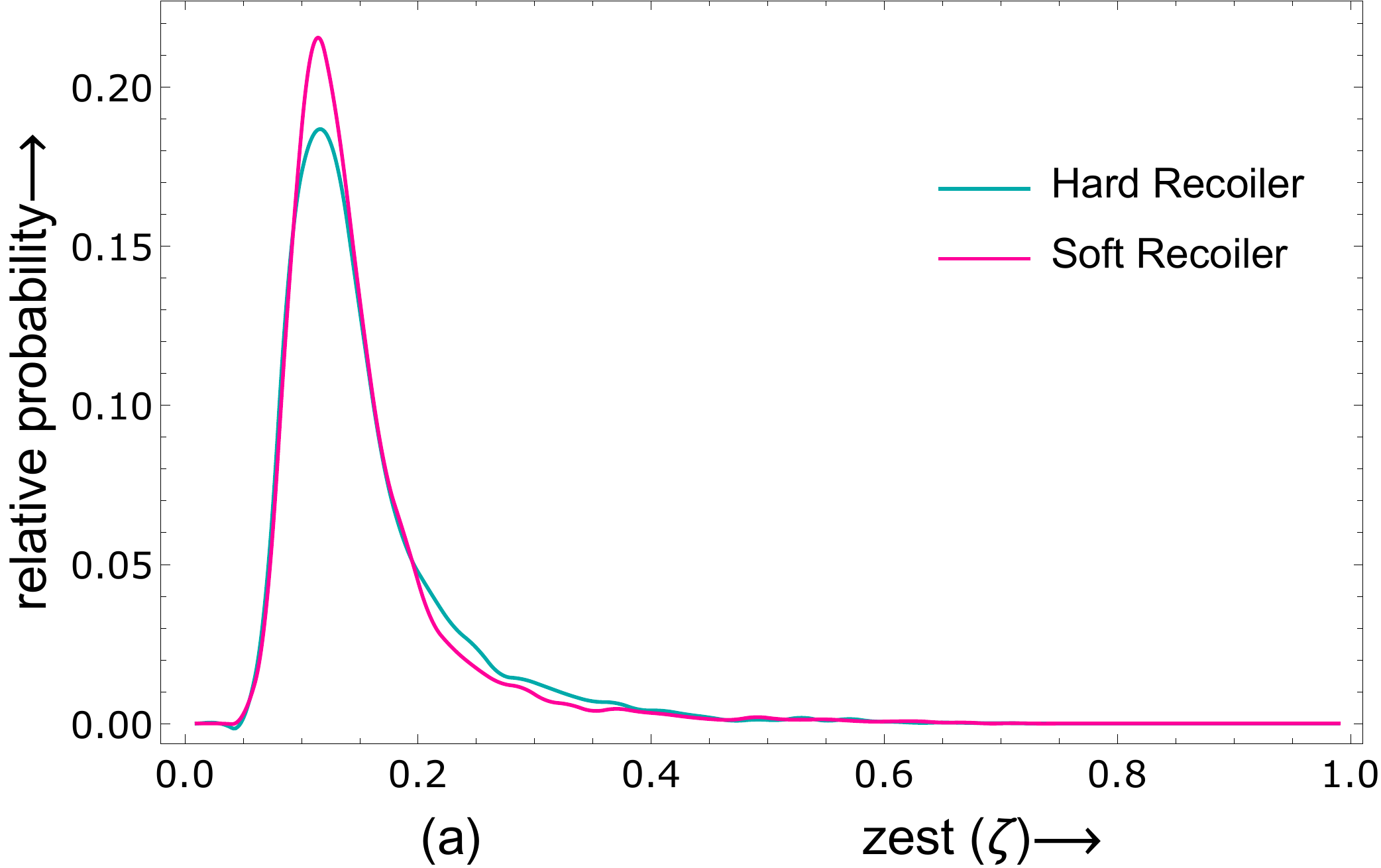}
  \includegraphics[width=0.43\textwidth]{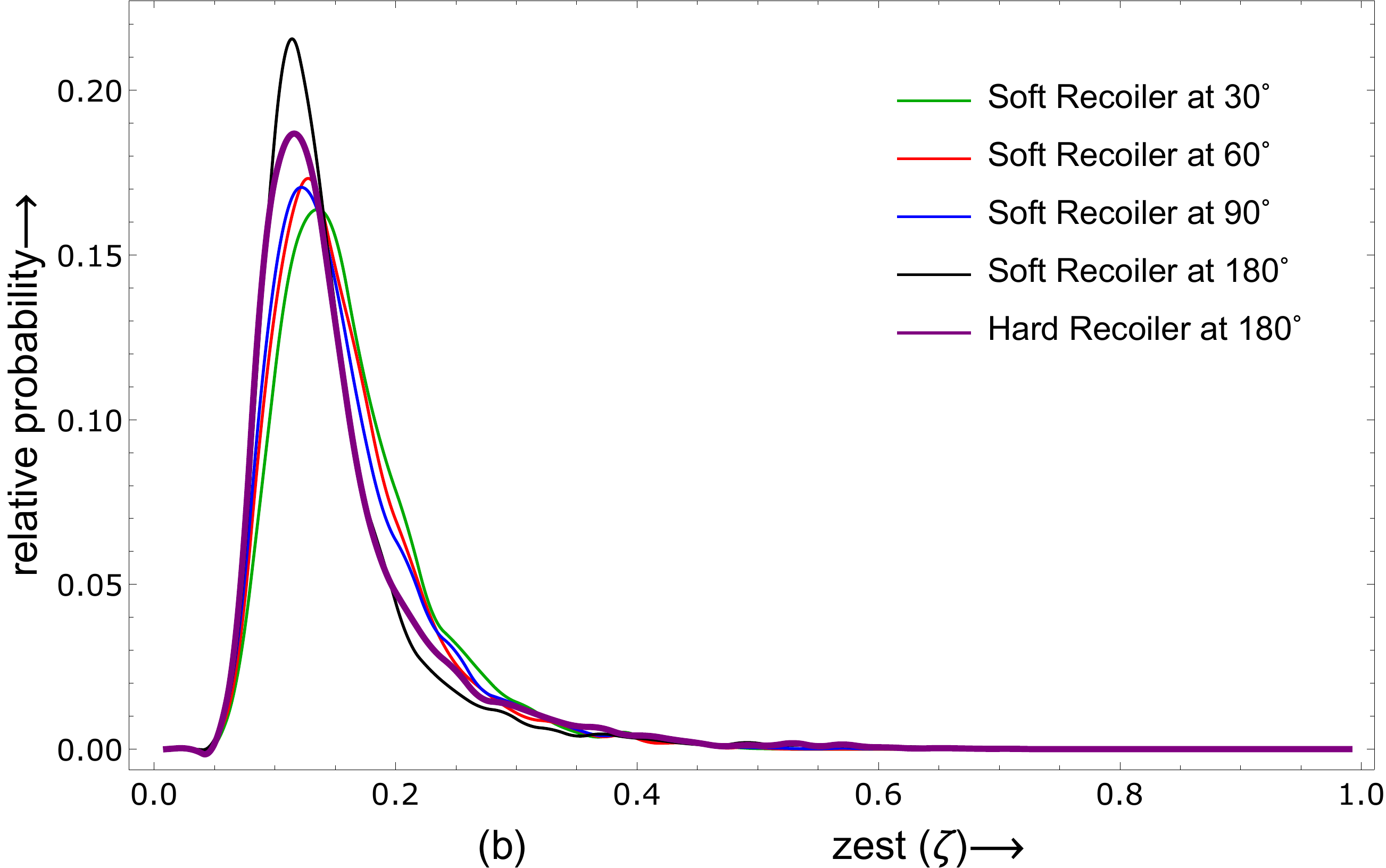}
 \caption{(a) Zest distribution of gluon-initiated jets with a hard gluon of energy about 300 GeV when the recoiling gluon is soft (blue) and hard (pink). (b) Zest distribution of gluon-initiated jets of energy about 300 GeV when the soft recoiler is taken at $30^{\circ}$ (green), $60^{\circ}$ (red), $90^{\circ}$ (light blue), $180^{\circ}$ (black) vs. a hard recoiler at $180^{\circ}$ (purple). } 
 \label{colorflow}
 \end{figure}
 
 \begin{figure}[t]
 \centering
  \includegraphics[width=0.44\textwidth]{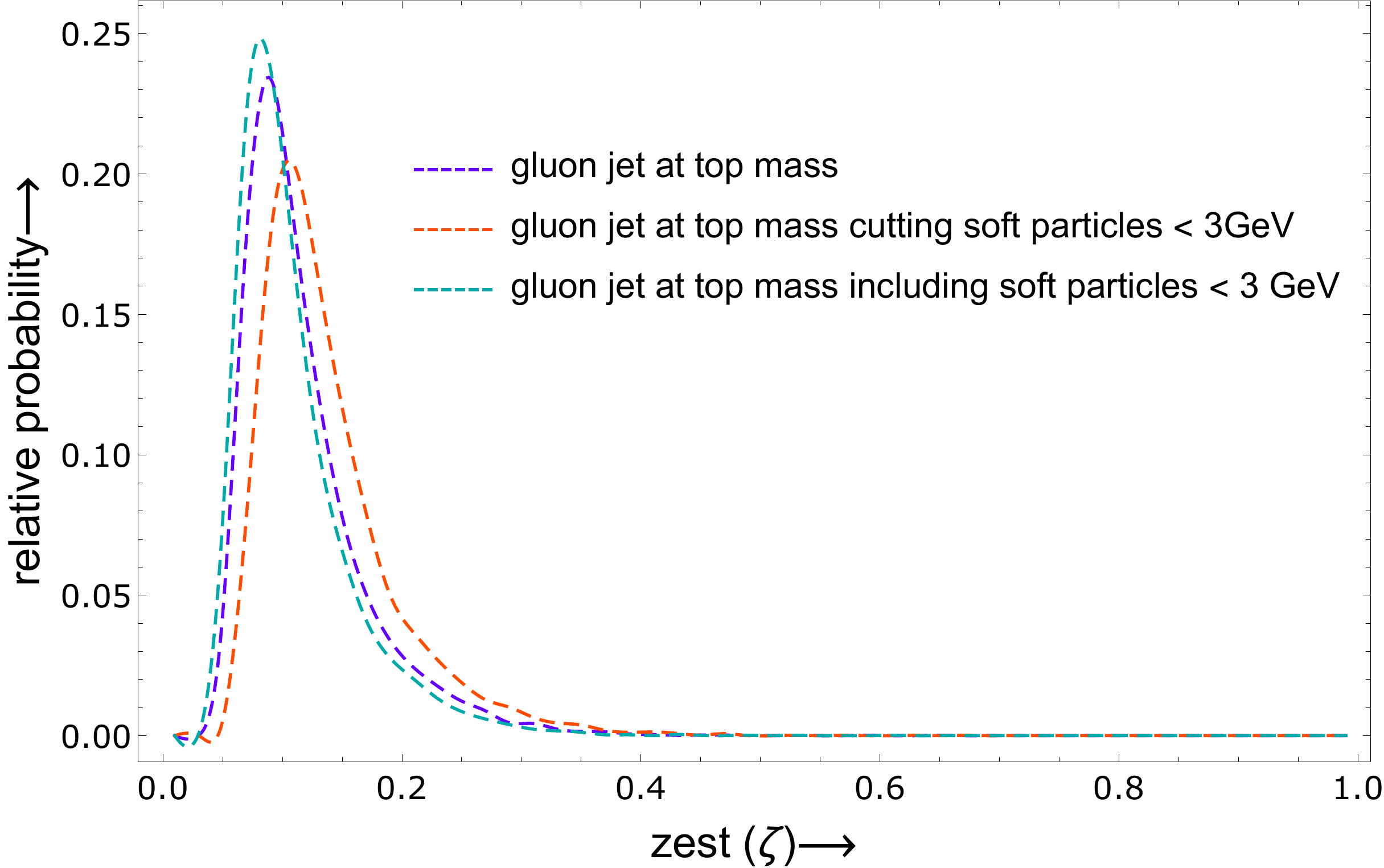}
 \caption{Zest distribution of gluon-initiated jets around top quark mass by including or excluding a few soft particles $< 3$ GeV to the jet.}
 \label{softphysics}
 \end{figure}

We also verify color flow independence of zest, discussed in the previous section, by looking at the zest-distribution of gluon jets recoiling against complementary color gluon at various angles. The result is shown in FIG.~\ref{colorflow}(a) and \ref{colorflow}(b). In FIG.~\ref{colorflow} we have considered jets of jet energy about 300 GeV, for two different cases: (i) a hard gluon of energy 300 GeV recoiling a soft gluon of energy 1 MeV at $180^\circ$, and (ii) a hard gluon of energy 300 GeV recoiling a hard gluon of energy 200 GeV at $180^\circ$. From the plot in FIG.~\ref{colorflow}(a), we note that the choice of the recoiling partner energy does not strongly effect the zest distribution. Furthermore, from  FIG.~\ref{colorflow} (b), we observe the color stability of gluon-initiated jets against the direction of its recoiling color neutralising partner. Since, the energy as well as the direction of the recoiling colored particle does not largely effect the analysis, for the rest of the paper, we will take the recoiling gluon as soft and back-to-back to the primary hard gluon.

As zest is largely insensitive to soft particles in the jet, its distribution curve is not effected by including or excluding a few soft particles in the jet. We verify this in FIG.~\ref{softphysics} where we have plotted the zest distribution curves for gluon-initiated jets of mass around the top quark mass with following three variations: (i) including all the particles in the jet, (ii) excluding all the soft particles less than $3$ GeV from the jet (iii) including few extra randomly generated soft particles less than $3$ GeV in the jet. We see that the zest distribution shifts only very slightly.

The stability of the gluon zest distribution against jet mass variation, color flow direction, and inclusion/exclusion of few soft particles makes it a potentially useful observable for vetoing gluon jets at the LHC.
For instance, a zest cut at about $\zeta \sim 0.3$ is expected to largely cut out the gluon background to heavy SM particles and hopefully also for heavier new physics resonances.

\section{Simulation Details}
\label{Simulation}

The details of our Monte Carlo simulations are presented below.

\begin{enumerate}

\item All jets are simulated using {\sc Pythia} 8 \cite{Sjostrand:2007gs} by inserting $W$ and $Z$ bosons with energy 500 GeV along the central axis (y-axis) and allowing them to decay to hadronic modes only.\footnote{For the study presented here, we have turned off all the strong, weak and electromagnetic decays of the primary hadrons formed from showering and subsequent hadronization. 
We do not study detector analysis in this paper.
} 

\item We collect all the particles in the forward hemisphere and call it a jet. Apart from hemisphere jets, sometimes we also construct anti-k$_t$ jets with different jet radii, R = \{0.5, 0.7, 0.9, 1.2\}, using the {\sc Fastjet} package~\cite{Cacciari:2011ma} in {\sc Pythia}.

\item For simulating a color-singlet top quark jet, we perform an $e^+\,e^-$ annihilation and allow the intermediate $Z$ boson to decay to a top-antitop pair at 500 GeV each, travelling back-to-back. We further constrain the top decay channels to hadrons only. 
Jets are constructed by dividing the event into two hemispheres using the thrust axis.

\item To shower a color singlet gluon-initiated jet, we insert an energetic gluon of energy 500 GeV and offshellness same as the heavy particle mass in the event record along the y-axis, while a soft onshell gluon of energy 1 MeV with opposite color is inserted in the opposite direction to the +y-axis. The direction of the color conserving gluon partner of the hard gluon does not effect the analysis, as zest is stable to global color flow of the partons (see FIG.~\ref{colorflow}). For convenience, we choose the direction of the recoiling gluon opposite to that of hard gluon.

\item The energetic gluon jet is identified by constructing the thrust axis and then taking all particles in the forward hemisphere.

\item The resulting gluon jets are accepted if their masses are within 10 GeV of the corresponding heavy particle mass. Although the energy of the jet does not effect zest, we observe that the selected gluons jets have energy ranging between 485 GeV to 500 GeV.

 \begin{figure*}
\centering
  \includegraphics[width=0.62\textwidth]{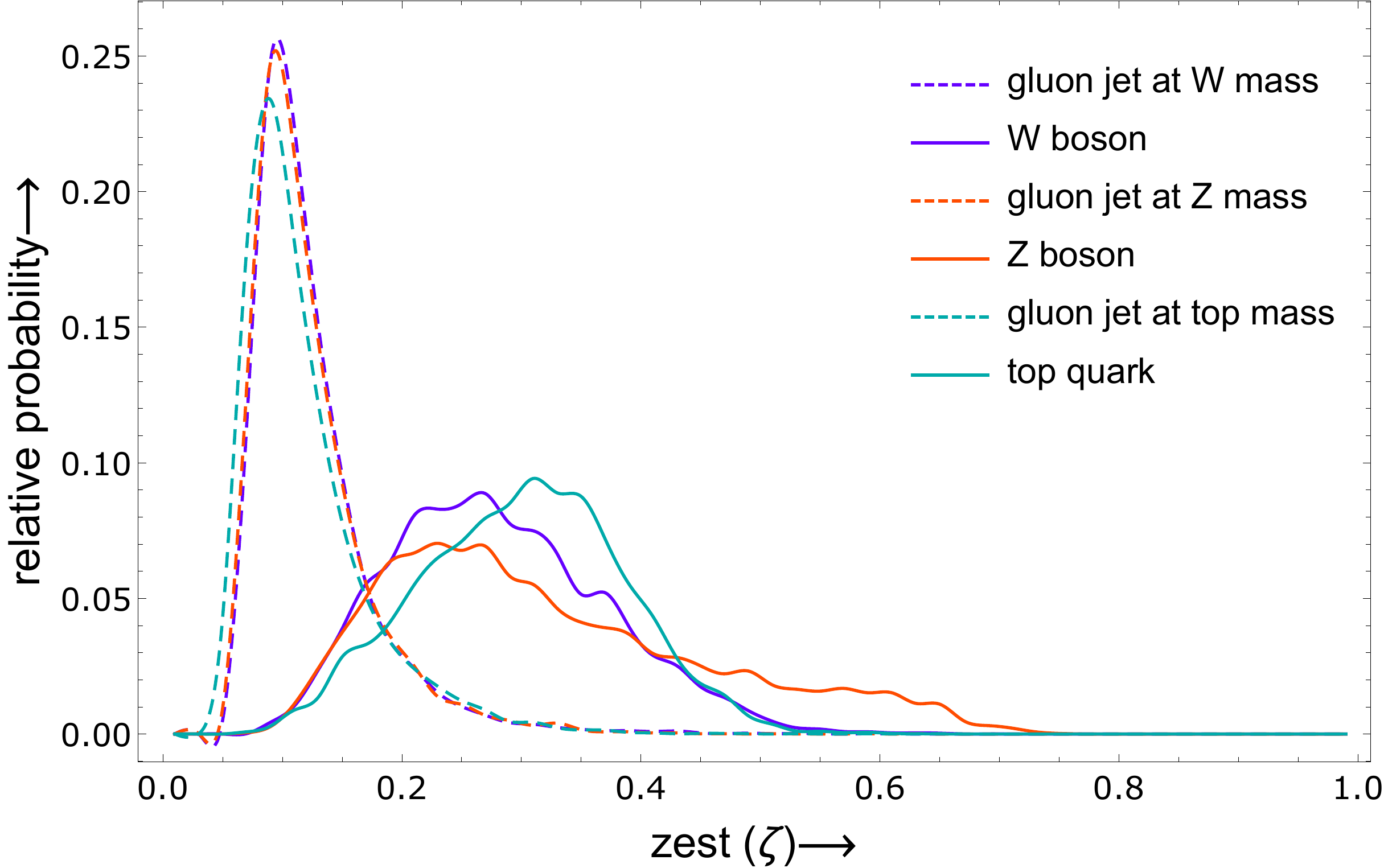}
  \caption{Zest distribution for $W,~Z$ boson and top quark initiated jets along with gluon jets of corresponding masses.}
 \label{zestplot}
 \end{figure*}

\item  In order
to test the sensitivity of our observable to non-perturbative physics, we also generate jets in two distinct ways: (a) by varying the default parameters of the hadronization model implemented in {\sc Pythia}, thereby producing a new set of final state events, and (b) by utilizing a different model for non-perturbative hadronization effects by using the {\sc Herwig}~\cite{Bellm:2015jjp} event generator. We study this non-perturbative physics dependence for $Z$ boson-initiated jet along with corresponding gluon background in Sec.~\ref{hadModel}.

\item The {\sc Herwig} event is sampled using the process $e^+\, e^- \to {\rm Z} \to {\rm jets}$ at a center of mass energy equal to the mass of the $Z$ boson, i.e., $E_{\rm cm} = m_Z$. We then boost all the final state particles along a direction, which we take as the +y-axis. The resulting boosted jet has a mass about the $Z$ boson and energy around 500 GeV.

\item In this work, we will not study Higgs-initiated jets because a Higgs boson dominantly decays to a $b\,\bar{b}$ pair which receives a large background from $Z\rightarrow b\,\bar{b}$ and $g\rightarrow b\,\bar{b}$, in which case the analysis proposed here does not offer a substantially good discrimination ability.
\end{enumerate}

\section{Zest as Filter}
\label{zestFilter}

As noted earlier, the zest distribution of gluon-initiated jets peaks at small values of the observable in comparison to their heavy particle counterparts. This accounts for an appreciable distinguishing criterion offered by zest in characterization of heavy particle jets from the background gluon jets.
We analyze this discriminating ability provided by zest by looking at the $\zeta$-distribution curves for $W,~Z$ boson and top quark-initiated jets along with the corresponding gluon-initiated jets with a mass about the heavy particle masses. This comparison is shown in FIG.~\ref{zestplot}. From the plot, we see that the zest distribution of the gluon-initiated jets peaks at relatively smaller values of $\zeta$ compared to their heavy particle counterparts. In addition, the zest distribution of gluon-initiated jets is found to be narrow, independent of jet energy and has a little overlap with a similar distribution of most other heavy particle jets. Moreover, we find that for the top quark-initiated jet, the zest distribution peaks at a slightly higher value than $W$ or $Z$ boson-initiated jets.

\begin{figure}
\centering
 \includegraphics[scale=0.35]{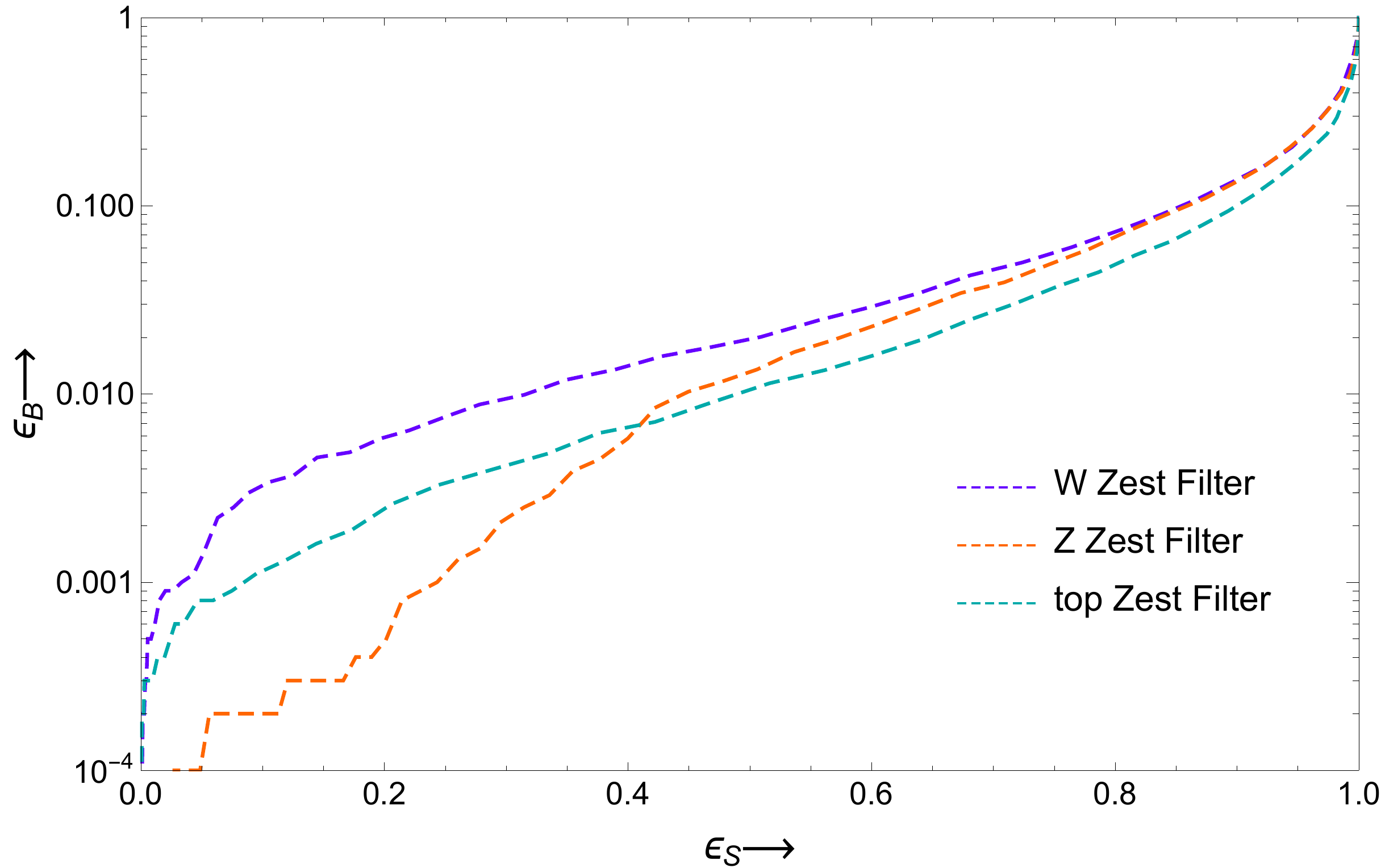}
  \caption{ROC curves for $W,~Z$ boson and top quark-initiated jets using {\it zest} as the binary classifier.}
 \label{ROC}
\end{figure}

The tagging efficiency of a zest-based binary classifier can be further evaluated by comparing the performance curves of these jets. In FIG.~\ref{ROC}, we present the relative operating characteristic (ROC) curves for the zest-based filter. We note that zest provides good signal statistics along with a high background rejection rate. For example, we note that for all the heavy particle jets considered, nearly 80 to 90\% of the signal stays in the accepted sample after applying a zest cut to remove 90\% of the gluons.

 \begin{figure}
\centering
  \includegraphics[width=0.47\textwidth]{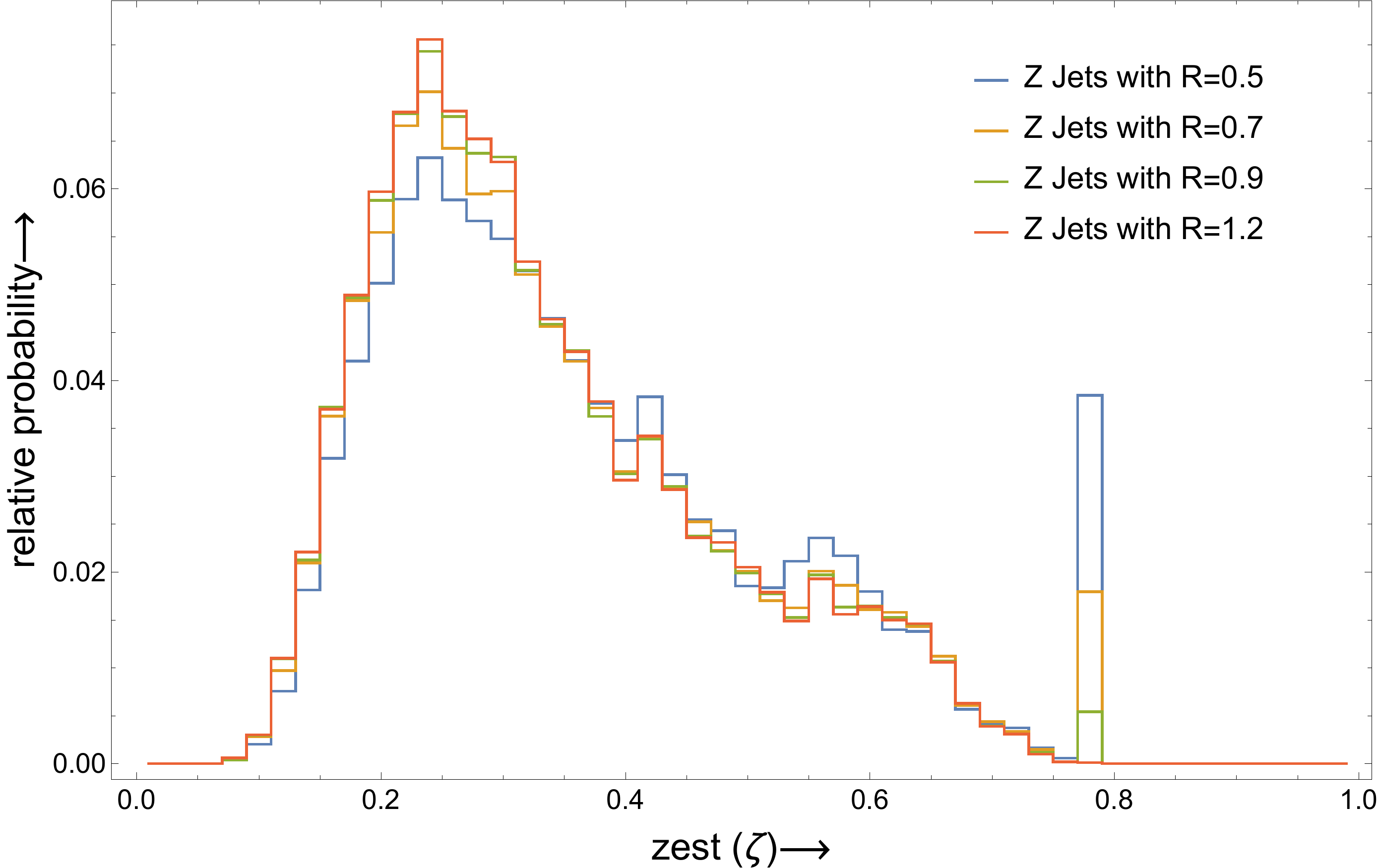}
  \caption{Zest distribution for $Z$ boson-initiated jets with different jet radii. Here jets are constructed using the anti-kT clustering algorithm for different $R$ values.}
 \label{ZJetsR}
 \end{figure}

\begin{figure*}
\centering
\includegraphics[width=0.7\textwidth]{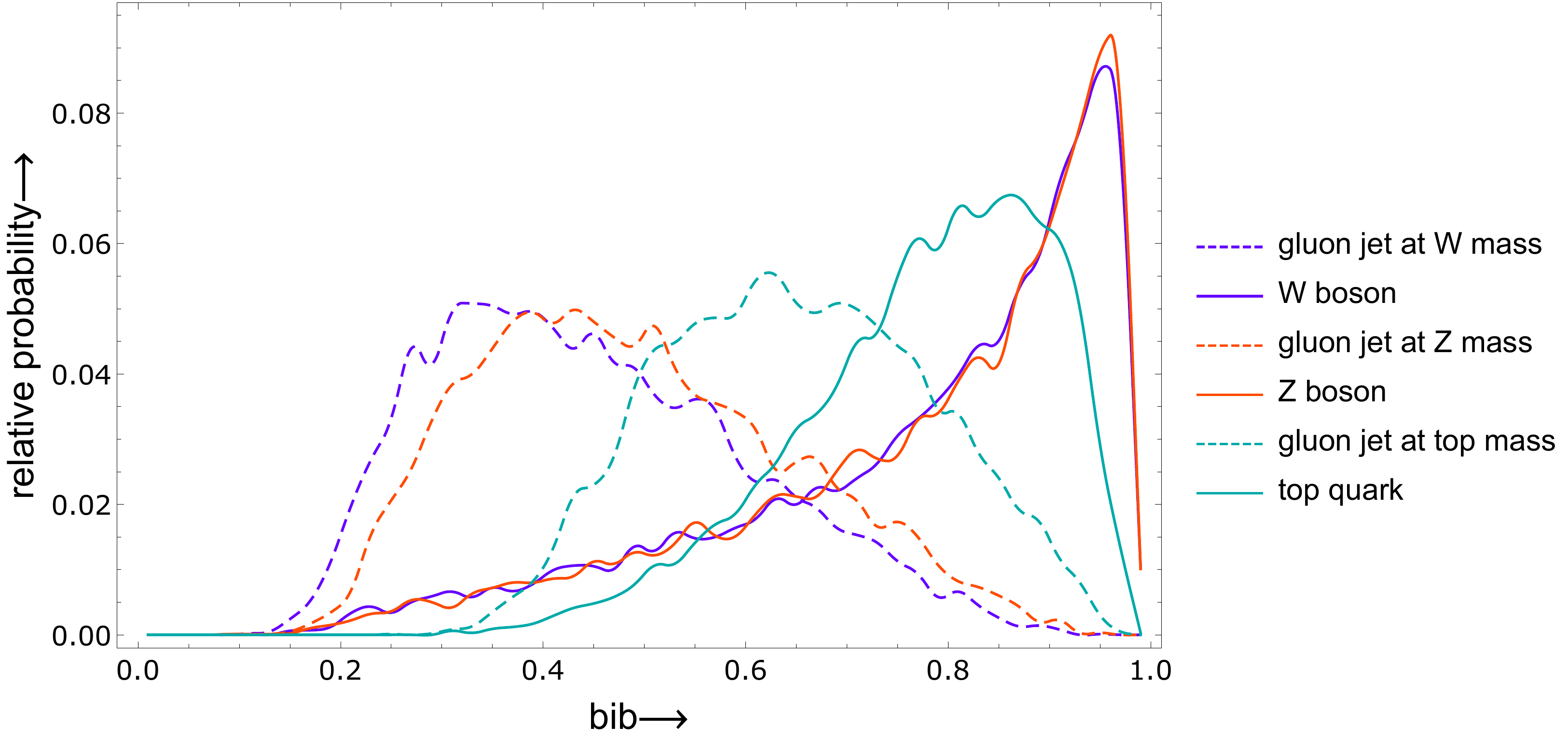}
\caption{$bib$-distribution of heavy particle initiated jets and gluon jets forming the background.} 
\label{bibplot}
\end{figure*}

In experiments, jets are often constructed by a suitable jet algorithm employing a cut on the jet cone radius $R$. At large jet radius, the wide angle soft particles contribute to the jet, while as the jet radius decreases, the observable receives contribution from collinear particles in the jet and a few soft particles. Since zest is not significantly effected by the inclusion or exclusion of a few soft particles, we anticipate the zest distribution to be stable against change in the jet cone radius. This is confirmed by FIG.~\ref{ZJetsR}, where $\zeta$-distribution curves with different jet radii, namely R = \{0.5, 0.7, 1.0, 1.2\}, are presented for $Z$-boson initiated jets. From the plot, we also note that the two particle peak at $\zeta \approx 0.77$ gets enhanced when the jet radius is sufficiently small, aligned with our expectation.
\section{Boost-invariant Broadening as Filter}
\label{bibFilter}

To contrast zest with an observable linearly dependent on transverse momentum $\vert {\bf{p}}_{T}\vert$ of the particles, we introduce a simple observable \textit{boost-invariant broadening} or $bib$ (for brevity) which is defined by the equation,
\begin{equation}
bib = \frac{1}{m_{J}} \sum_{i\epsilon Jet} \vert {\bf p}_{T i} \vert = \frac{P_T}{m_J}\,.
\label{bib}
\end{equation}
Here as well, all transverse momenta are measured with respect to the jet axis and $P_T$ is the same quantity as defined in the context of zest. It is interesting to note that $bib$ is similar to the event shape observable \textit{broadening}~\cite{Dokshitzer:1998kz}, however the use of $ m_{J}$ instead of $E_{J}$ renders it invariant under boosts made along the jet axis. Zest and $bib$ have boost-invariance along the jet axis and jet energy independence in common with each other. Although there can be other observables which can be chosen for a bi-variate analysis, $bib$ provides a simple contrast to zest : (i) $bib$ is a linear function of the transverse momentum of the particles while zest is not, (ii) $bib$ is calculable using standard perturbative techniques\footnote{As $bib$ is defined by the ratio of two IRC safe observables, it is essentially Sudakov safe~\cite{Larkoski:2013paa, Larkoski:2015lea} which implies that even though the observable does not has a valid fixed order expansion in $\alpha_s$, it can still be calculated using all-order resummation. This ensures that the singular region of the phase space for the observable is exponentially suppressed.}, while zest is collinear unsafe and relies on a suitable Monte Carlo generator that encodes a definite model to capture the hadronization effects for its computation, and (iii) $bib$ is sensitive to the global color flow of partons while zest is insensitive to it. The results presented in this section and for the  bi-variate analysis in the next section, are strictly for the case in which the soft gluon is back-to-back with the primary hard gluon.   

\begin{figure}
\centering
\includegraphics[width=0.47\textwidth]{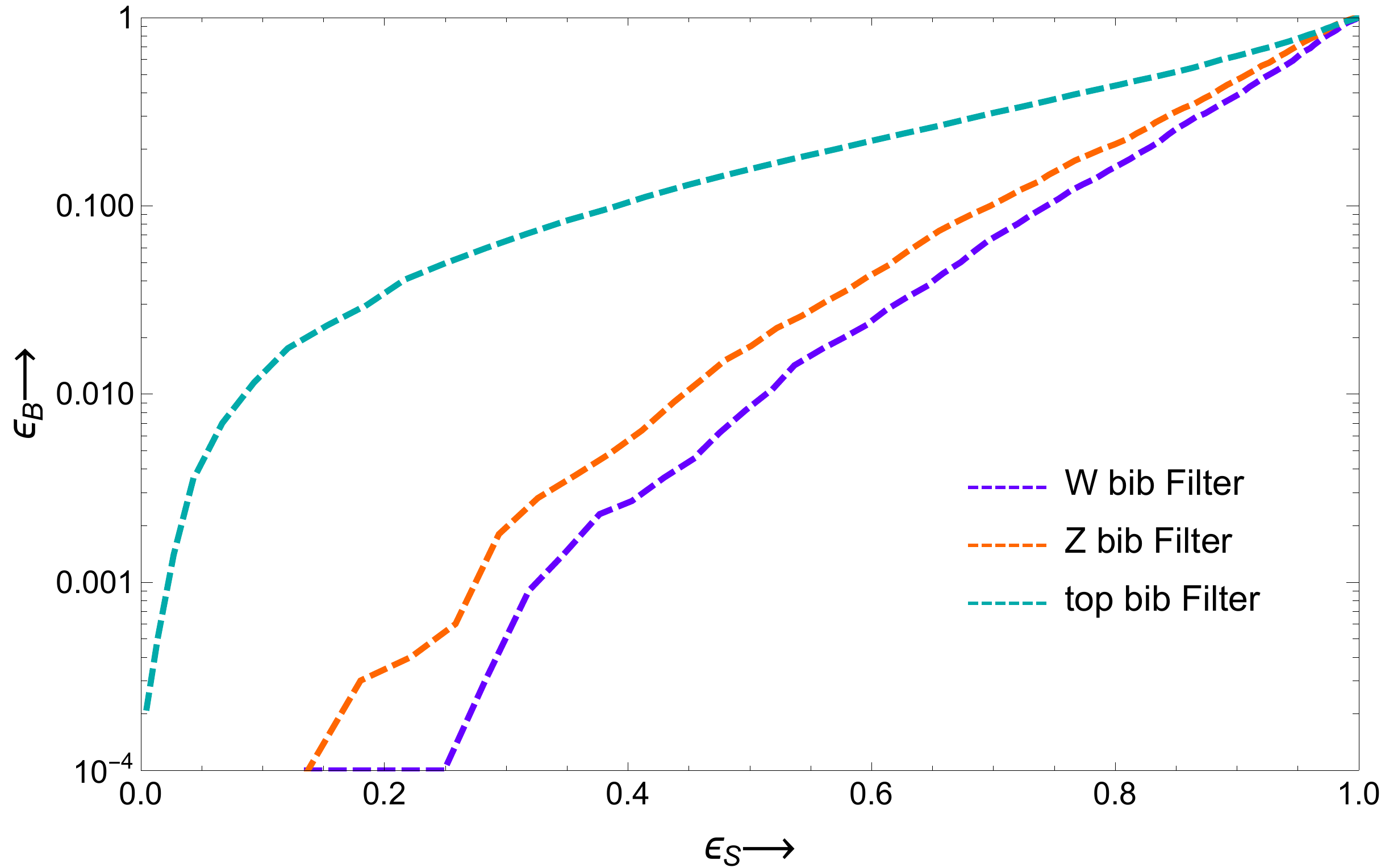}
\caption{ROC curves for heavy particle jets with $bib$ as the classifier.}
\label{bibROC}
\end{figure}

The physics of the primary splitting of heavy particles leading to partonic decays is different from that of branching of quarks and gluons. Parton splittings dictated by Quantum Chromodynamics (QCD) are dominated by soft and collinear radiation having small transverse momenta, thus we expect gluon initiated jets to have a peak at a relatively smaller value in $bib$-distribution compared to jets initiated by heavy particles at the same jet mass.
The comparison of $bib$-distributions for heavy particles and corresponding gluon jets is shown in FIG.~\ref{bibplot}.
We also present the ROC curves for $bib$ when used as a binary classifier in FIG.~\ref{bibROC}. From the plot, we find that $bib$ acts as an appreciable classifier for $W,~Z$ boson-initiated jets but does not offer good discrimination for heavier top quark-initiated jets. This is in contrast with the zest ROC curves of FIG.~\ref{ROC}. Overall, zest provided better classifier performance compared to $bib$. Nonetheless, $bib$ also contrasts zest in the stability of gluon distributions with respect to the jet mass. We have also verified that the two observables do not have a significant linear correlation. Thus, we expect that the two observables in conjunction can be used to improve the signal to background ratio. We present this bi-variate analysis in the next section.

\section{Bi-variate Analysis}
\label{bivariate}

\begin{figure}
\centering
 \includegraphics[width=0.48\textwidth]{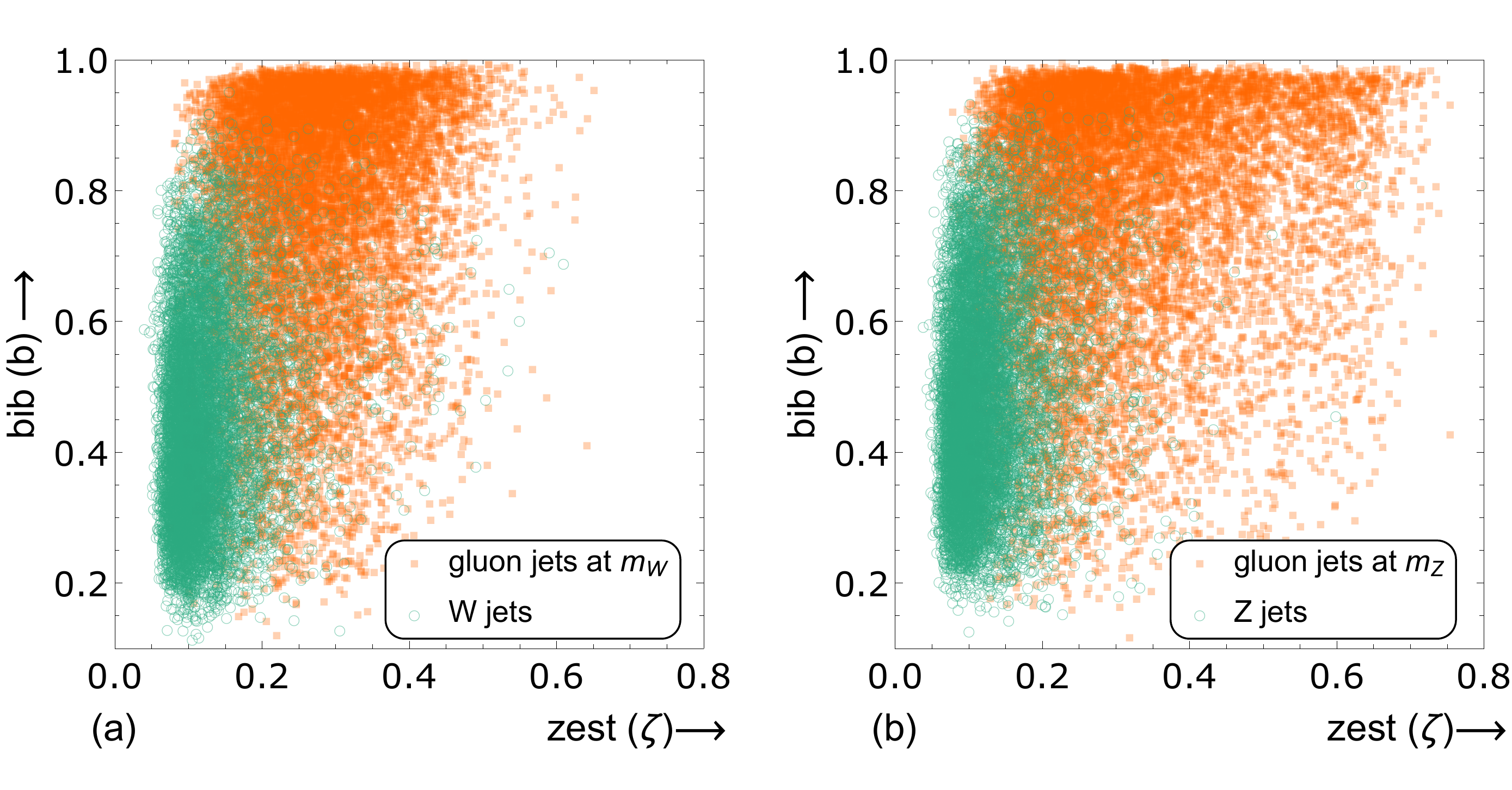}\\
  \includegraphics[width=0.35\textwidth]{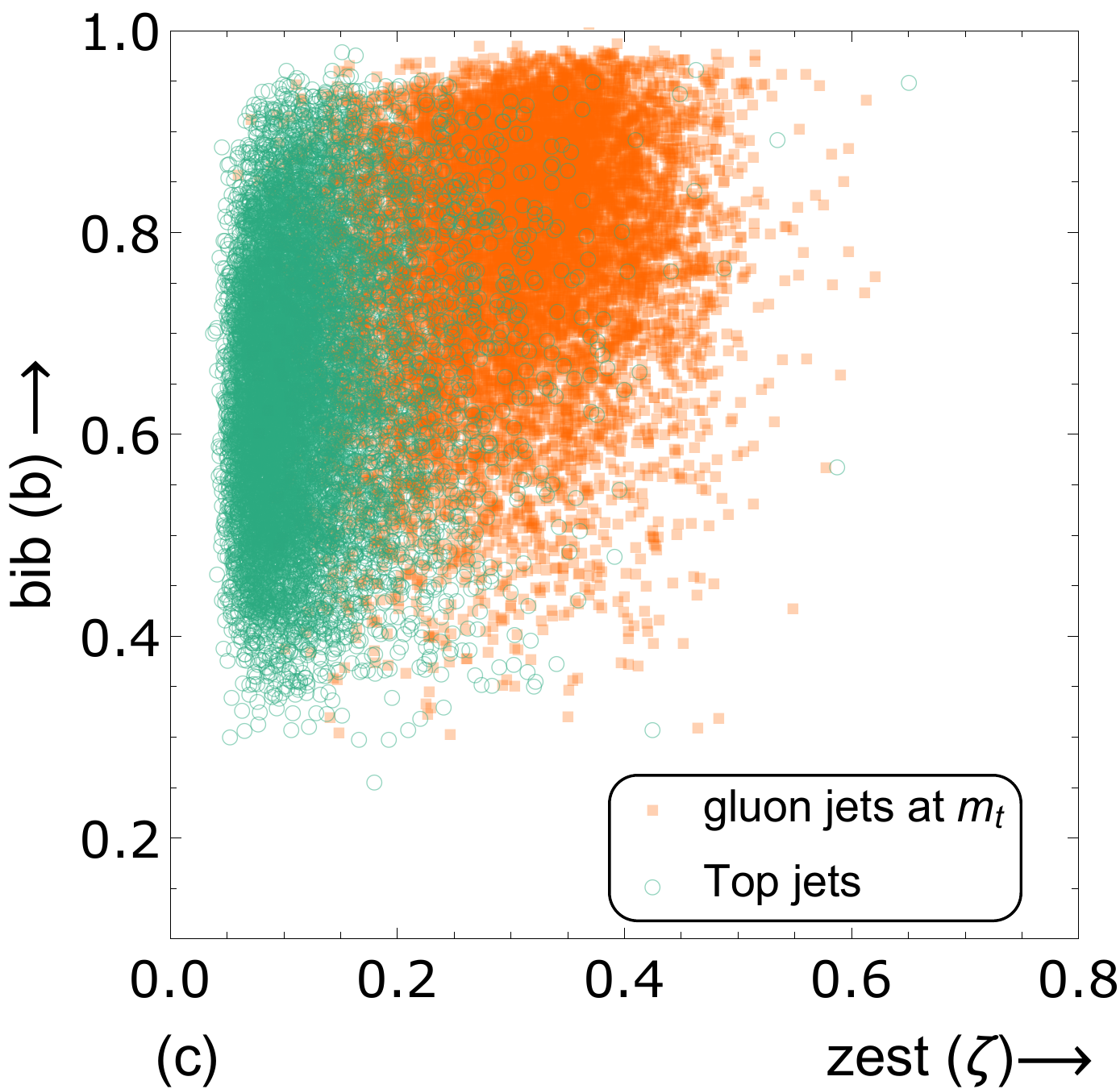}
 \caption{(a)-(c) Scatter plots in $bib$-zest plane for heavy particle and corresponding gluon jets.}
 \label{scatter}
\end{figure}

Discrimination for the originating particle can be further improved with a multivariate analysis. Recently, multivariate approaches are becoming extremely popular as they are expected to provide a more detailed characterization of the QCD radiation pattern within a jet which can be exploited to further enhance the new physics searches at the LHC~\cite{Larkoski:2017jix, Marzani:2019hun}. For the bi-variate analysis proposed here, we generate scatter plots in the \textit{bib}-zest plane for gluon initiated jets and heavy particle jets, as shown in FIG.~\ref{scatter} (a)-(c).  
We note that, in general, gluon jets occupy the extreme left corner of the $bib$-zest plane, while heavy particle jets are mostly concentrated towards the upper right region in the plot, thus providing statistical discrimination between the two types of jets.

\begin{figure}
\centering
 \includegraphics[width=0.48\textwidth]{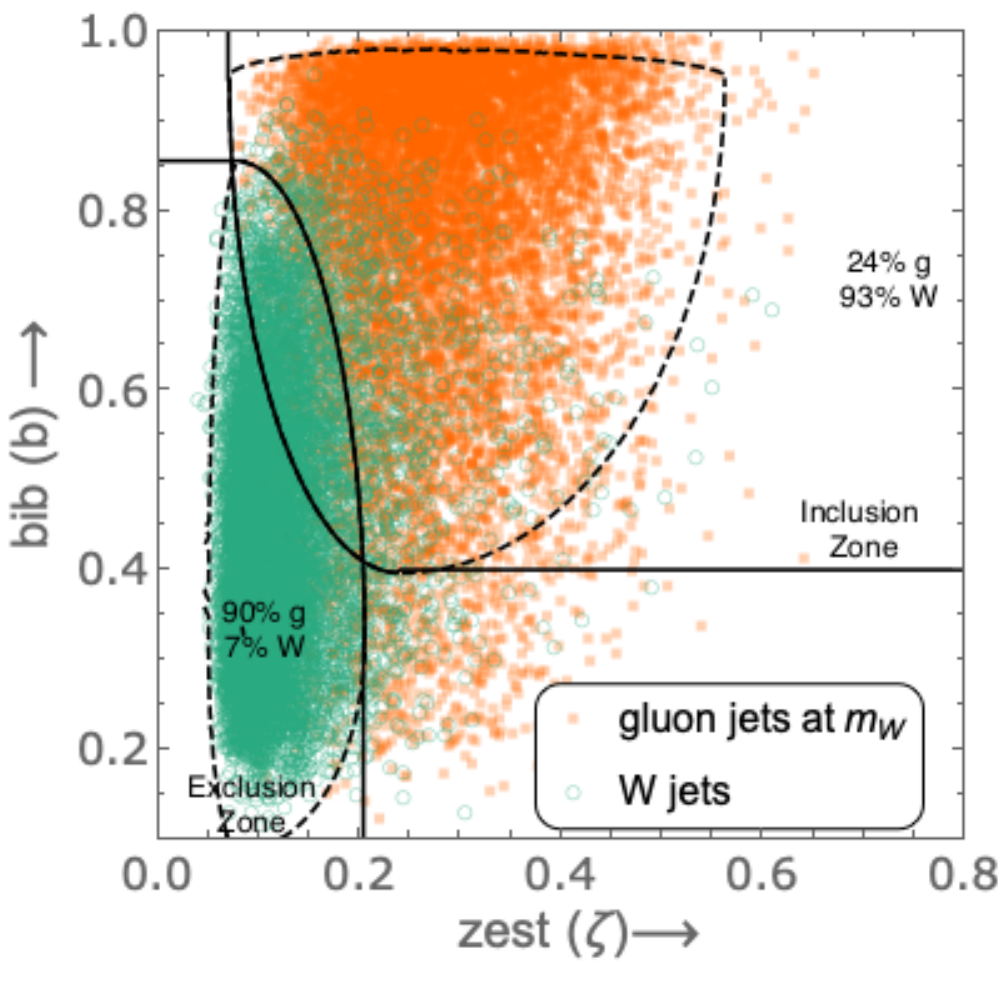}
 \caption{Scatter plot in $bib$-zest plane for $W$-boson jet and the corresponding gluon jet, with the 90\% Gaussian contour and inclusion and exclusion-zone cuts specified on it.}
 \label{Wscatter}
\end{figure}

\begin{figure*}
\centering
 \includegraphics[width=0.7\textwidth]{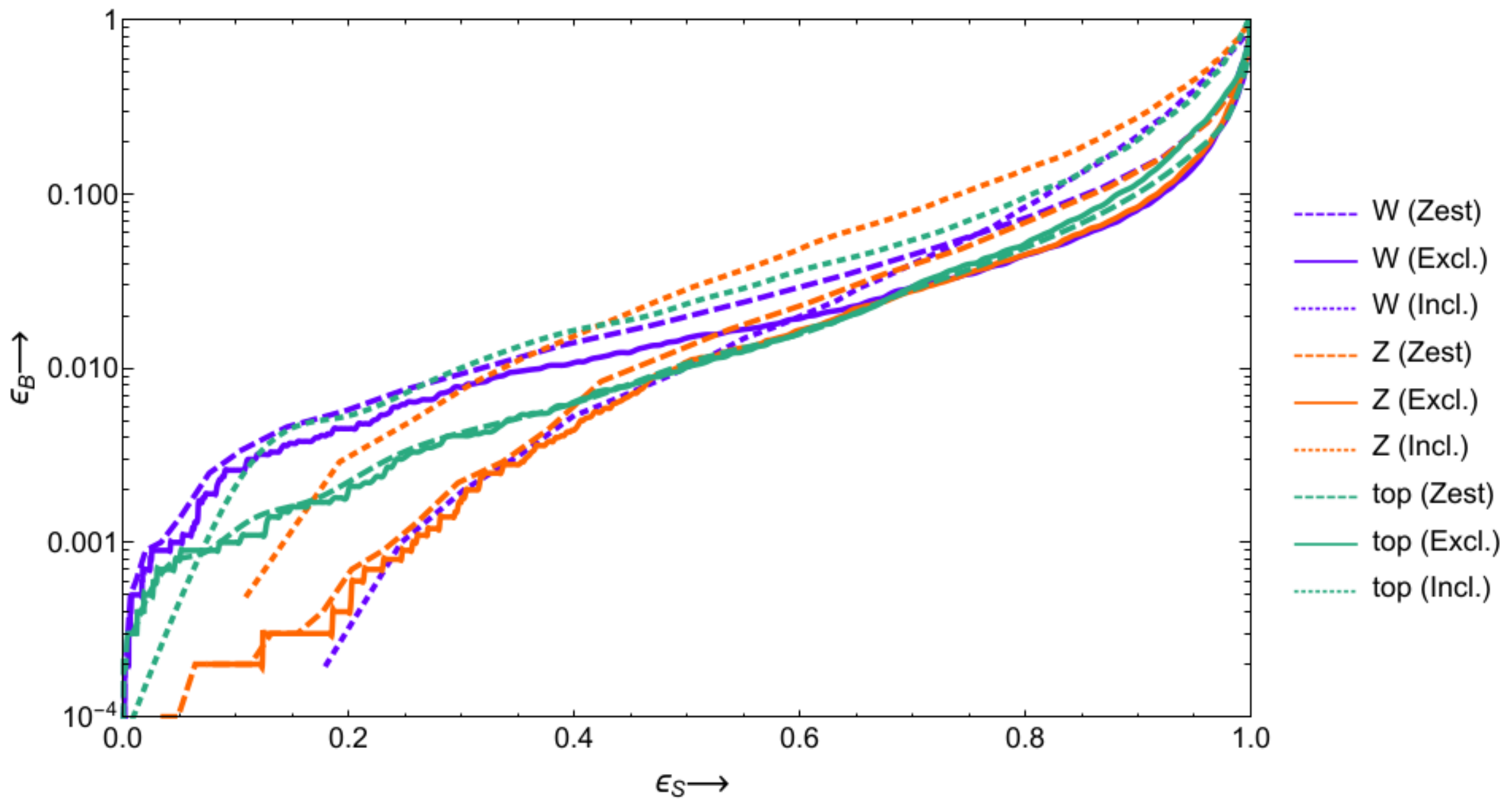}
 \caption{ROC curves for heavy particle jets with zest, inclusion and exclusion$-$filter based binary classifiers.}
 \label{allRoC}
\end{figure*}

The zest-$bib$ distributions shown in FIG.~\ref{scatter} are reasonably well-approximated by an asymmetric two-dimensional Gaussian distribution. We fit both the heavy particle and gluon jet distributions with an asymmetric Gaussian distribution. On the fitted Gaussian distributions, we draw contours of different heights. An example for the 90\% contour, illustrated with a dashed line, is shown in FIG.~\ref{Wscatter} for the $W$-boson initiated jets along with the corresponding gluon background. By 90\% contour, we refer to the closed curve that includes 90\% of the jets inside it. The tangents are drawn parallel to $bib$-axis and zest-axis on the gluon distribution contour to intercept with the plot frame at the bottom and to the left, respectively. The solid line represents an open curve made by combining the tangents and right-side of the contour meeting the tangents. The background region is defined as the area to the left of the solid curve. Similar exercise is also done on the heavy particle initiated jet distribution. The signal region is defined as the region to the right of the solid curve on the heavy particle distribution. We define two new cuts (1) the  `inclusion-zone' cut and (2) the `exclusion-zone' cut. The `inclusion-zone' cut is the solid curve on the signal while the `exclusion-zone' cut is the solid curve on the background. We note that for the case of a $W$ boson-initiated jet with 90\% of the signal as the desired value, an exclusion-zone cut provides better statistics over an inclusion-zone cut.

The relative performance of these bi-variate cuts in comparison to zest alone is studied through ROC curves in FIG.~\ref{allRoC}. We present three types of ROC curves for each type of heavy particle jet (a) zest based cuts only (represented by dashed lines), (b) cuts based on inclusion-zone obtained by including the percentage of events contained within the primary contour (represented by dotted lines), and (c) cuts based on exclusion-zone obtained by excluding the percentage of events contained within the primary contour (represented by solid lines). From the plot, we note that for a high signal rate, exclusion-zone statistics provides slightly better discrimination than only zest-based cuts for the $W,~Z$ bosons, however inclusion/exclusion-zone provide no significant improvement in comparison to a zest only filter for the top quark-initiated jet.
On the other hand, inclusion-zone statistics is more efficient when high gluon rejection is required and small signal rate is acceptable.

\section{Generalized Zest ($p$-zest)}
\label{pzest}
As pointed out earlier in Sec.~\ref{Zest}, in the special case of $n$-equal transverse momenta particles in the jet, zest is a function of {\it particle multiplicity} in a jet. In this section, we aim to generalize the observable such that the particle multiplicity emerges as a limiting case of generalized zest. We define the generalization of zest (or, `$p$-zest' for brevity) as follows:

\begin{equation}
\zeta_p = -\frac{1}{\log\Big(\sum_i\, e^{-P_T^{(p)}/\vert {\bf p}_{T i}^{p}\vert}\Big)}\, .
\end{equation}
Here the sum is over all the final state particles in the jet, as earlier, and $P_{T}^{(p)}$ is defined as
\begin{equation}
P_T^{(p)} = \sum_i \vert {\bf p}_{T\,i}^{\,p} \vert \, ,
\end{equation}
where $p > 0$. The ability to tune the parameter $p$ may also provide a new way of looking at the jet substructure by allowing us to vary the net contribution of the soft sector.

For $p > 1$, the contribution of the soft particles to the determination of $P_T^{(p)}$ will reduce. 
Thus, mostly collinear or energetic particles will contribute to $p$-zest.

 \begin{figure}
 \centering
  \includegraphics[width=0.475\textwidth]{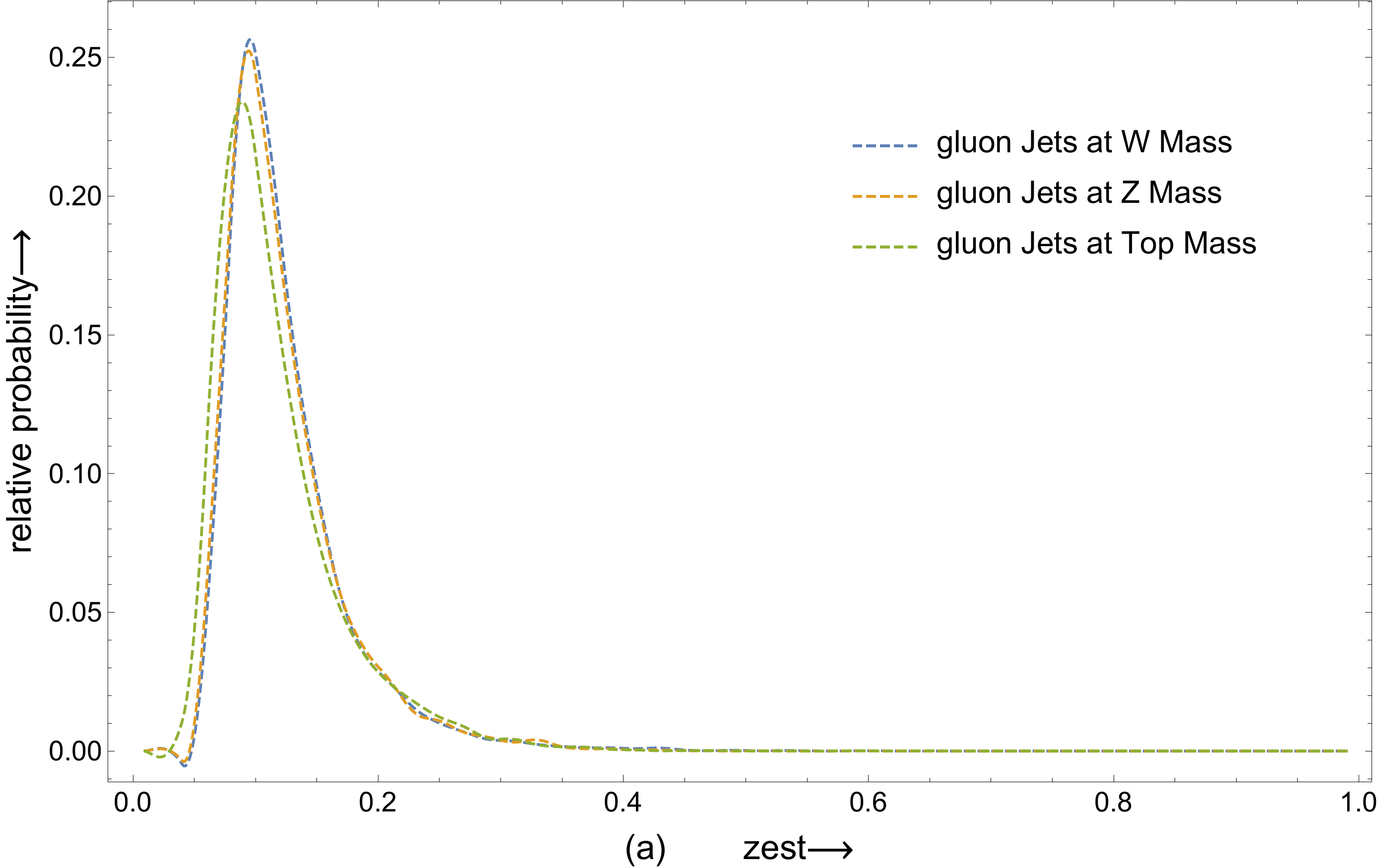}
  \includegraphics[width=0.48\textwidth]{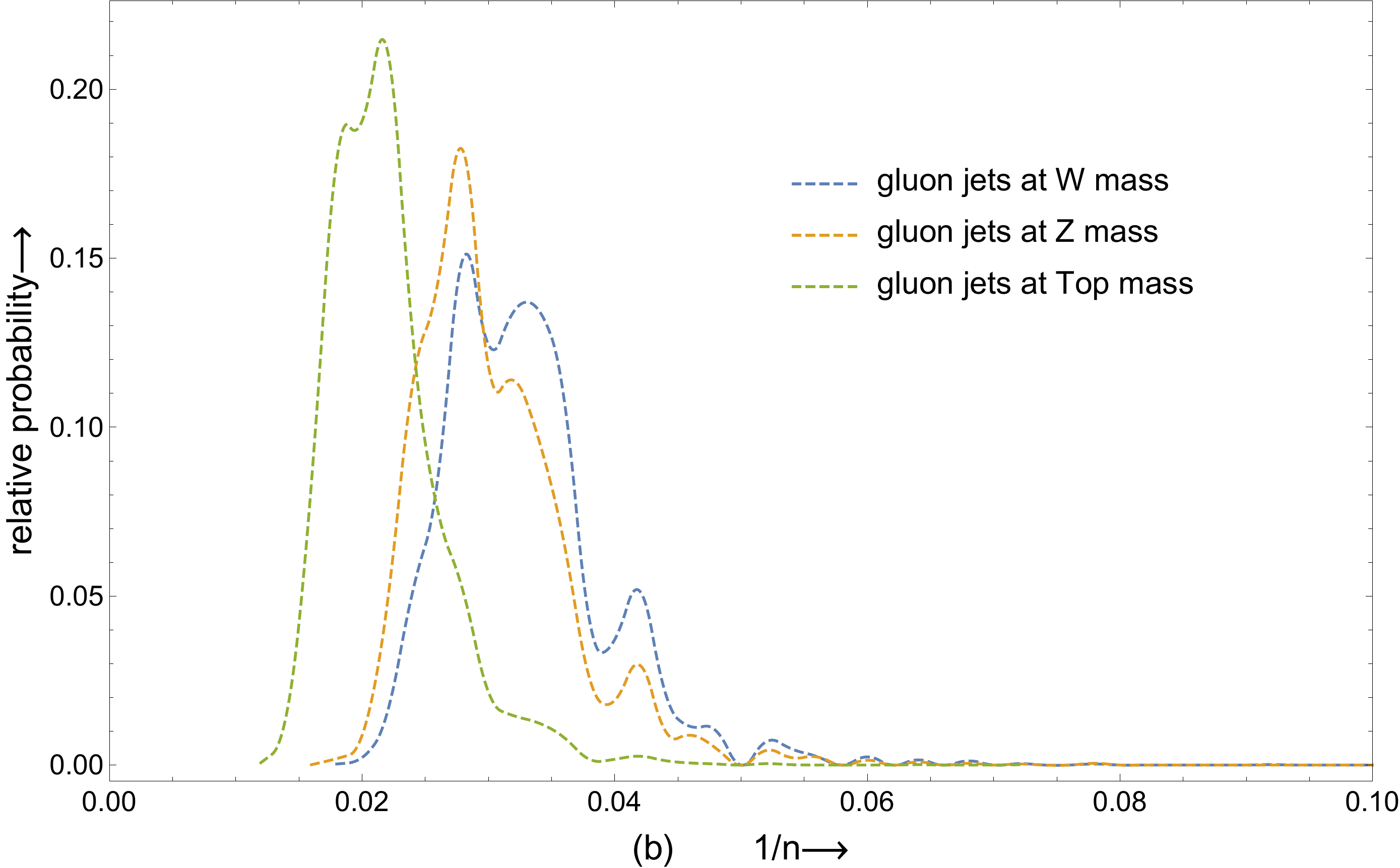}
 \caption{(a)Zest distribution of gluon jets at various heavy particle mass windows, and (b) $1/n$ distribution of gluon jets at various heavy particle mass windows.}
 \label{zvsmult_Mass}
 \end{figure}

\begin{figure}[t]
 \centering
  \includegraphics[width=0.48\textwidth]{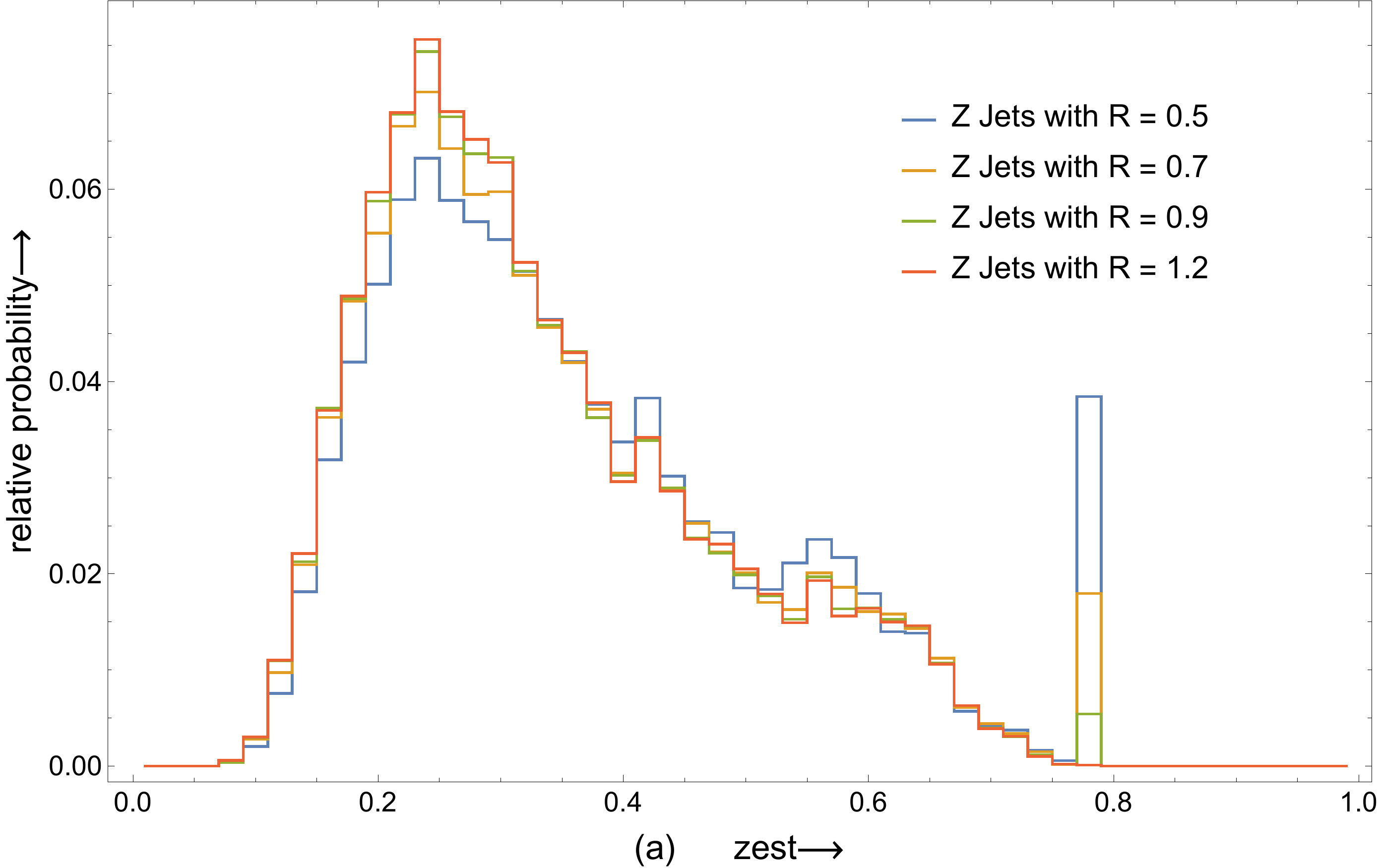}
  \includegraphics[width=0.47\textwidth]{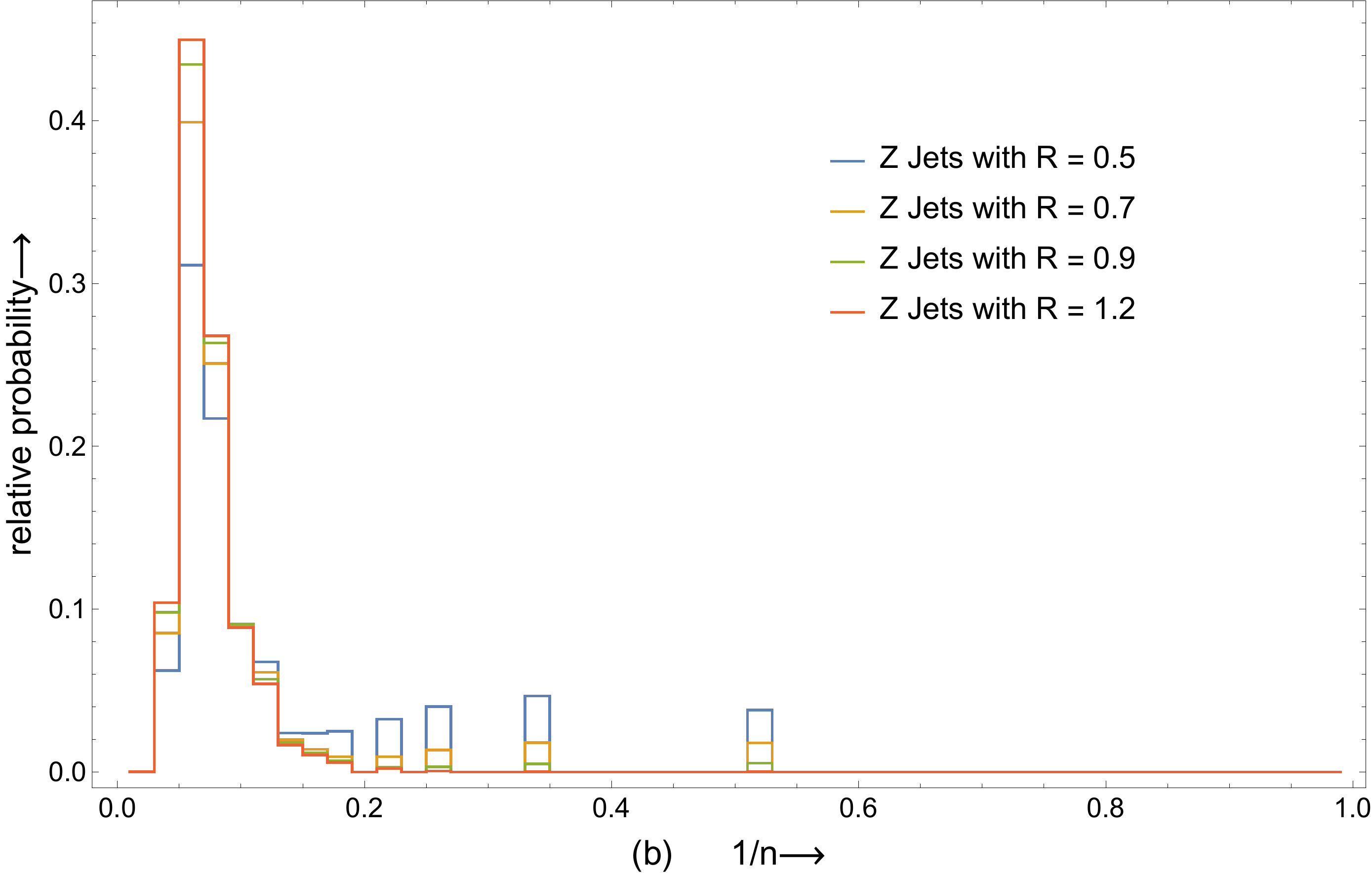}
 \caption{(a)Zest distribution of $Z$ boson-initiated jets for different $R$ values, and (b) $1/n$ distribution of $Z$ boson-initiated jets for varying jet radii.}
 \label{zvsmult_R}
 \end{figure}

In the extreme limit of $p \to \infty$, only the leading parton contributes and $\zeta_{\infty}$ approaches 1. Therefore, this limit offers no useful jet discrimination ability.   
In contrast, in the limit $p \to 0$, all particles contribute equally giving
\begin{equation}
\zeta_0 = \frac{1}{n-\log n} \, ,
\label{eq:mult}
\end{equation}
where $n$ is the number of particles in the jet. For a large enough $n$ such that $n \gg \log n$, the above expression reduces simply to the inverse of particle multiplicity.  
Thus, we note that closer the value of $p$ is to 0, the stronger is the correlation of $p$-zest and multiplicity. As $p$ is increased, the correlation becomes smaller and vanishes for the extreme limit of $p \to \infty$.
Though the observables zest (with $p=1$) and multiplicity are correlated,  
zest exhibits interesting properties such as stability against change in the jet mass and stability against change in the jet radius for gluon-initiated jets, which are absent in the $1/n$ distribution. This is confirmed by the plots in FIG.~\ref{zvsmult_Mass} and FIG.~\ref{zvsmult_R}. In FIG.~\ref{zvsmult_Mass}(a) and~\ref{zvsmult_Mass}(b), the zest and the $1/n$ distribution curves for gluon-initiated jets with masses about heavy particle mass, are presented.   
From the plots, we see that the $1/n$ distribution shifts to the left as the mass of the jet is increased while the zest distribution remains more or less unchanged. \footnote{Note that this is not to say that $1/n$ does not offer gluon jet vetoing ability, instead it may very well provide the ability to veto gluon jets like zest. From the trend observed in FIG.~\ref{zvsmult_Mass}(b), we expect the heavier jets to shift further to the left and 
hence a cut at around 0.04 may cut out the gluon jets largely. However, as we will show in Sec.~\ref{top}, for the heavy top quark-initiated jets, zest offers an appreciable improvement over multiplicity.}
Similarly, the stability of the zest distribution curves to varying the jet radius is shown in FIG.~\ref{zvsmult_R}(a) for a $Z$ boson-initiated jet of jet energy 300 GeV while the corresponding $1/n$ curves are shown in FIG.~\ref{zvsmult_R}(b)\footnote{Here we have used jets of initial energy 300 GeV to demonstrate this difference. This is because a $Z$ boson at 500 GeV is extremely collimated and $R = 0.5$ is sufficient to capture all the radiations originating from this extremely boosted $Z$ boson.}. From the plots, we clearly see that the zest distribution curves are mostly stable to the change in jet radius while for the $1/n$ distribution, the curve for $R=0.5$ appears to be shifted away from other $R$ values. This is expected as zest is mostly uneffected by the inclusion or exclusion of a few soft particles while the multiplicity curves are somewhat sensitive to it.

Noting these differences, we investigate the discrimination ability offered by $p$-zest over particle multiplicity in a jet and zest ($p$=1). It will be interesting to find $p$ that provides the best discrimination. For this,  we consider two type of heavy particle jets : (a) a $Z$ boson-initiated jet, and (b) a top quark-initiated jet. 
The detailed analysis of these results is presented in the following subsections.

\subsection{$Z$ boson-initiated jet}
\label{Z}
\begin{figure}
\centering
 \includegraphics[width=0.5\textwidth]{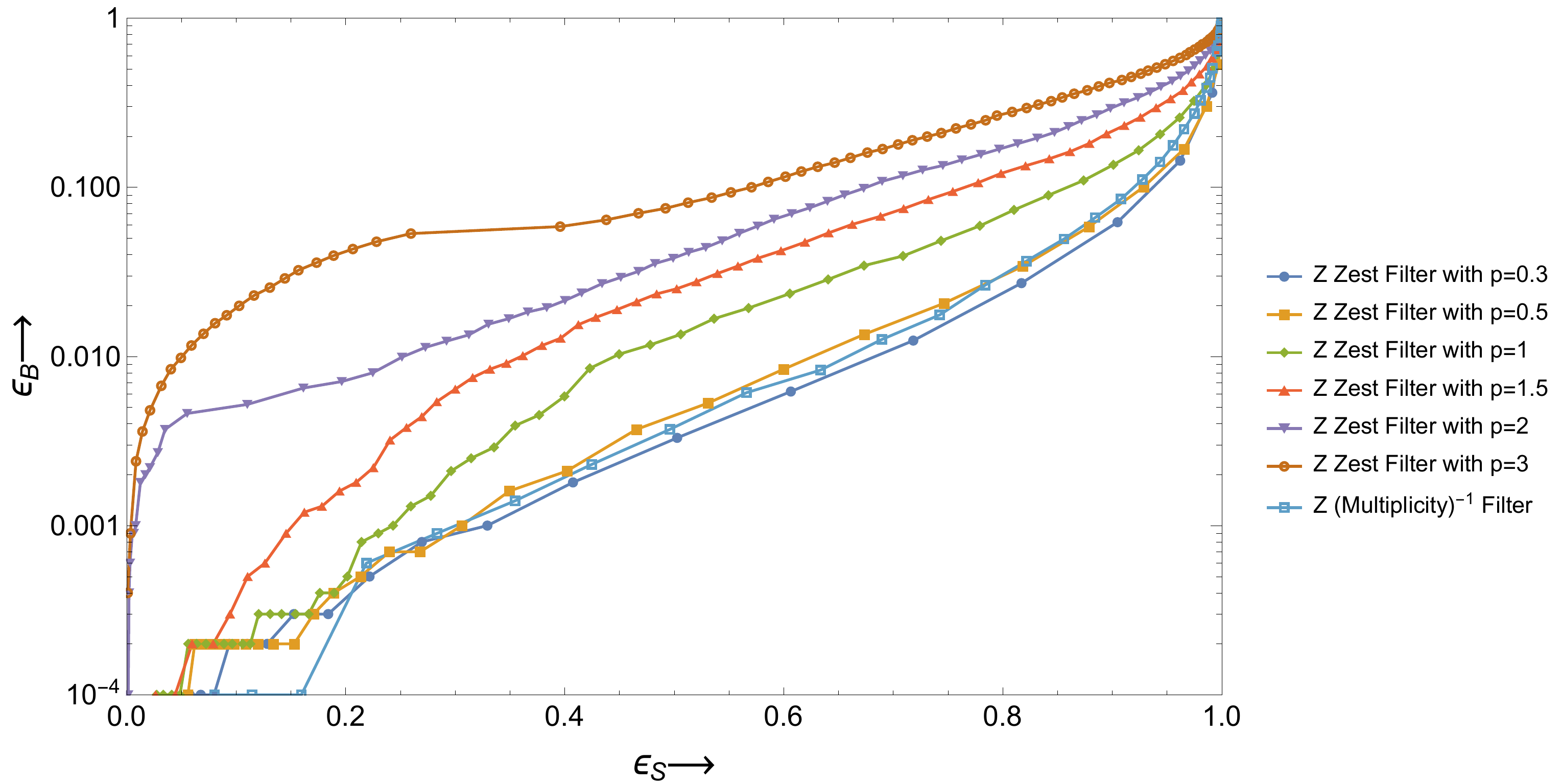}\\
\caption{ROC curves for $Z$ boson-initiated jets with various choices of $p$ compared against the inverse of multiplicity as a filter.}
 \label{ZFilter}
\end{figure}

\begin{figure}
\centering
\includegraphics[width=0.47\textwidth]{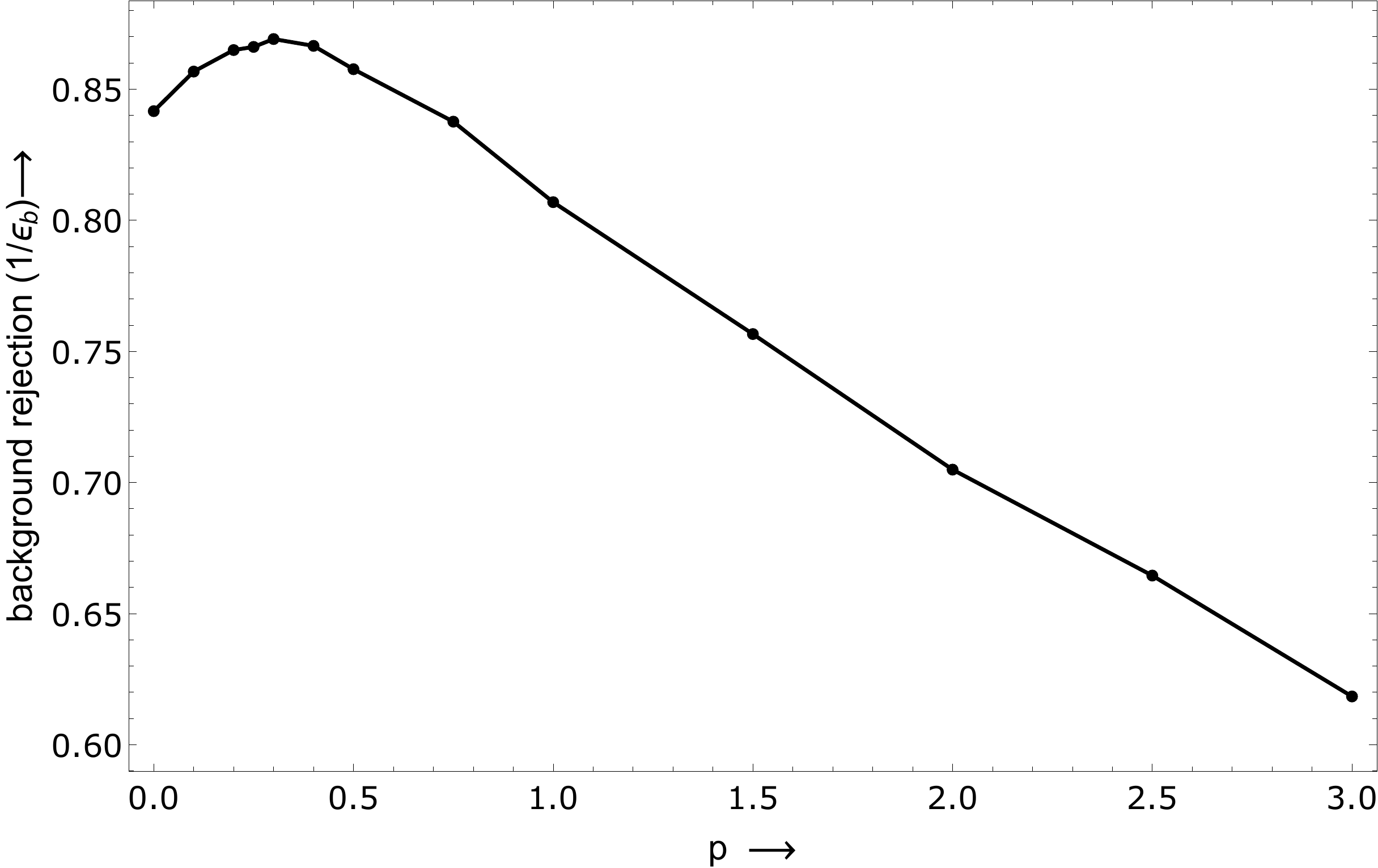}
\caption{Discrimination power offered by $p$-zest at fixed signal efficiency of 80\% for $Z$-boson initiated jets.}
 \label{ZFilter1}
\end{figure}

For the $Z$ boson-initiated jet, we generate the ROC curves for $p \!=\!\{0.3, 0.5, 1, 1.5, 2, 3\}$ and particle multiplicity, as shown in FIG.~\ref{ZFilter}. The solid cyan line with open squares on it illustrates the multiplicity distribution while the other solid lines with the corresponding legends represent the $p$-zest distributions. From the plots, we find that multiplicity provides a better discrimination in comparison to zest ($p$ = 1) while $p$-zest with $p = 0.3-0.5$ provides the optimal discrimination.
This is also confirmed by looking at FIG.~\ref{ZFilter1} where with 80\% of signal efficiency as the accepted value, we plot the curve for the background jet rejection offered by $p$-zest with varying values of parameter $p$. From the plot, we find that $p=0.3-0.5$ provides the best discriminating ability for a $Z$-boson initiated jet, although it is not significantly better than just multiplicity.

\subsection{top quark-initiated jet}
\label{top}

\begin{figure}
\centering
 \includegraphics[width=0.5\textwidth]{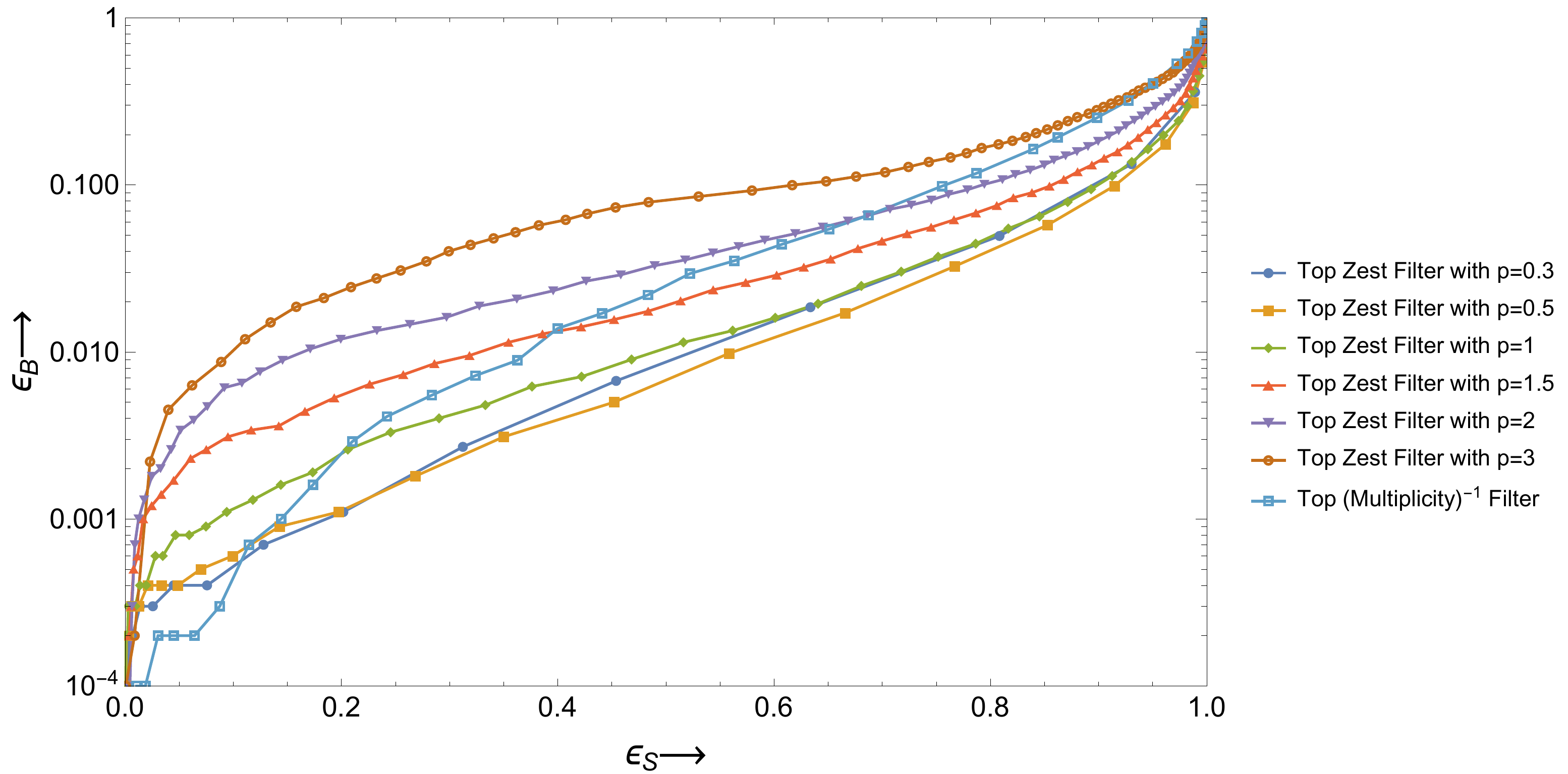}
\caption{ROC curves for top quark-initiated jets with various choices of $p$ compared against the inverse of multiplicity as a filter.}
 \label{TopFilter}
\end{figure}

\begin{figure}
\centering
 \includegraphics[width=0.47\textwidth]{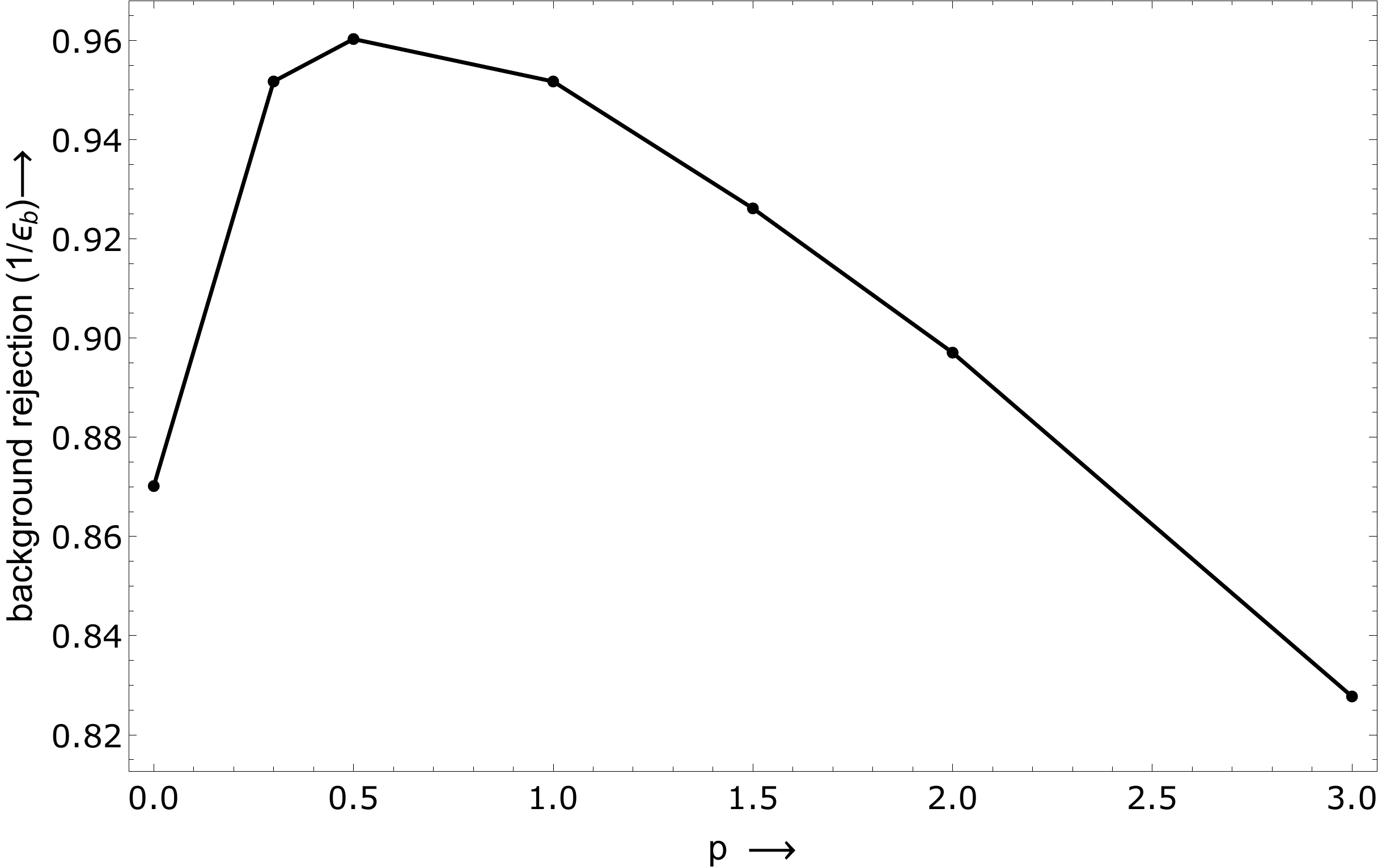}
\caption{Discrimination power offered by $p$-zest at fixed signal efficiency of 80\% for top quark initiated jets.}
 \label{TopDiscriminate}
\end{figure}

The ROC curves for top-initiated jets with $p$-zest and $1/n$ as the observables are presented in FIG.~\ref{TopFilter}. Interestingly, here we find that $p$-zest for $p= 0.3 - 1.0$ performs significantly better than the $1/n$ distribution. The same is confirmed further by the background rejection efficiency curve with respect to $p$ at fixed signal efficiency of 80\%, as shown in FIG.~\ref{TopDiscriminate}.

From this analysis, we find that $p$-zest allows us to find the $p$ value that optimizes the discrimination. We note that a $p$ value of about $0.3$ to $0.5$ is optimal for the heavy particles studied here.

It is interesting to note that the discrimination ability offered by our preliminary $p$-zest study for the top quark-initiated jets is comparable with that of a class of ML-based top taggers ~\cite{Kasieczka:2019dbj}. Although our study does not include proton-proton collisions and detector effects considered in~\cite{Kasieczka:2019dbj}, we are not claiming a strict comparison of our results with ML-based techniques. While these ML algorithms offer the highest discrimination ability for heavy particle jets, the particular features that offer this improvement remain largely unknown. $p$-Zest may offer some insight into the physics of ML-based taggers. Therefore, we propose that studying IRC unsafe observables based on physical principles may provide a window into the physics hidden in ML based techniques.

\section{Hadronization Model dependence}
\label{hadModel}

As pointed out earlier, $p$-zest is collinear unsafe and cannot be calculated by standard perturbative techniques, therefore we rely on Monte Carlo event generators that incorporate a suitable hadronization model for its computation. In this section, we study the effect of varying hadronization models on zest. This will allow us to understand how the observed value of zest is sensitive to the modeling of non-perturbative effects. This check is important as the observable can only be computed for hadronic final states.

\begin{figure*}
\centering
 \includegraphics[width=0.7\textwidth]{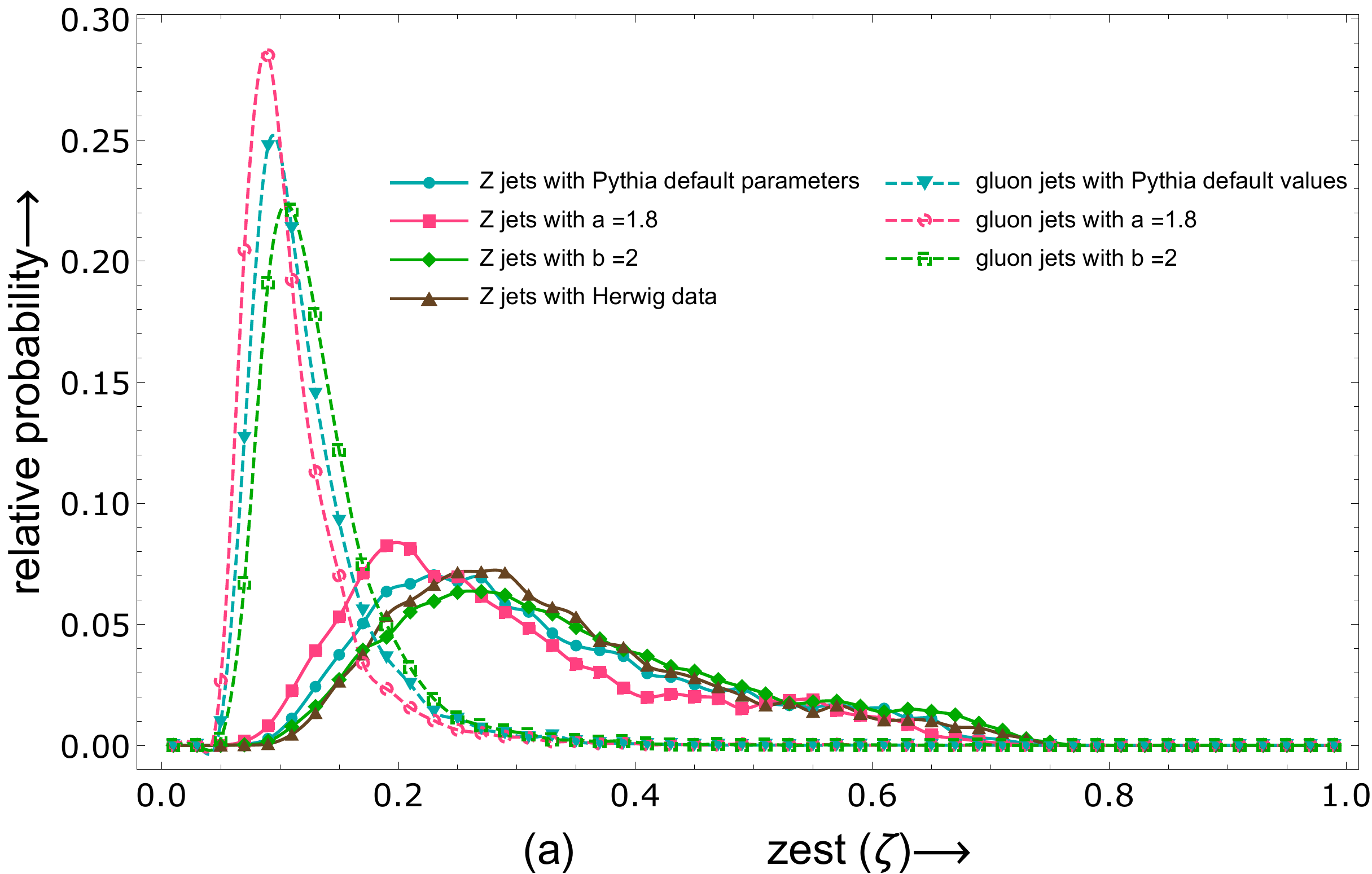}\\
 \includegraphics[width=0.7\textwidth]{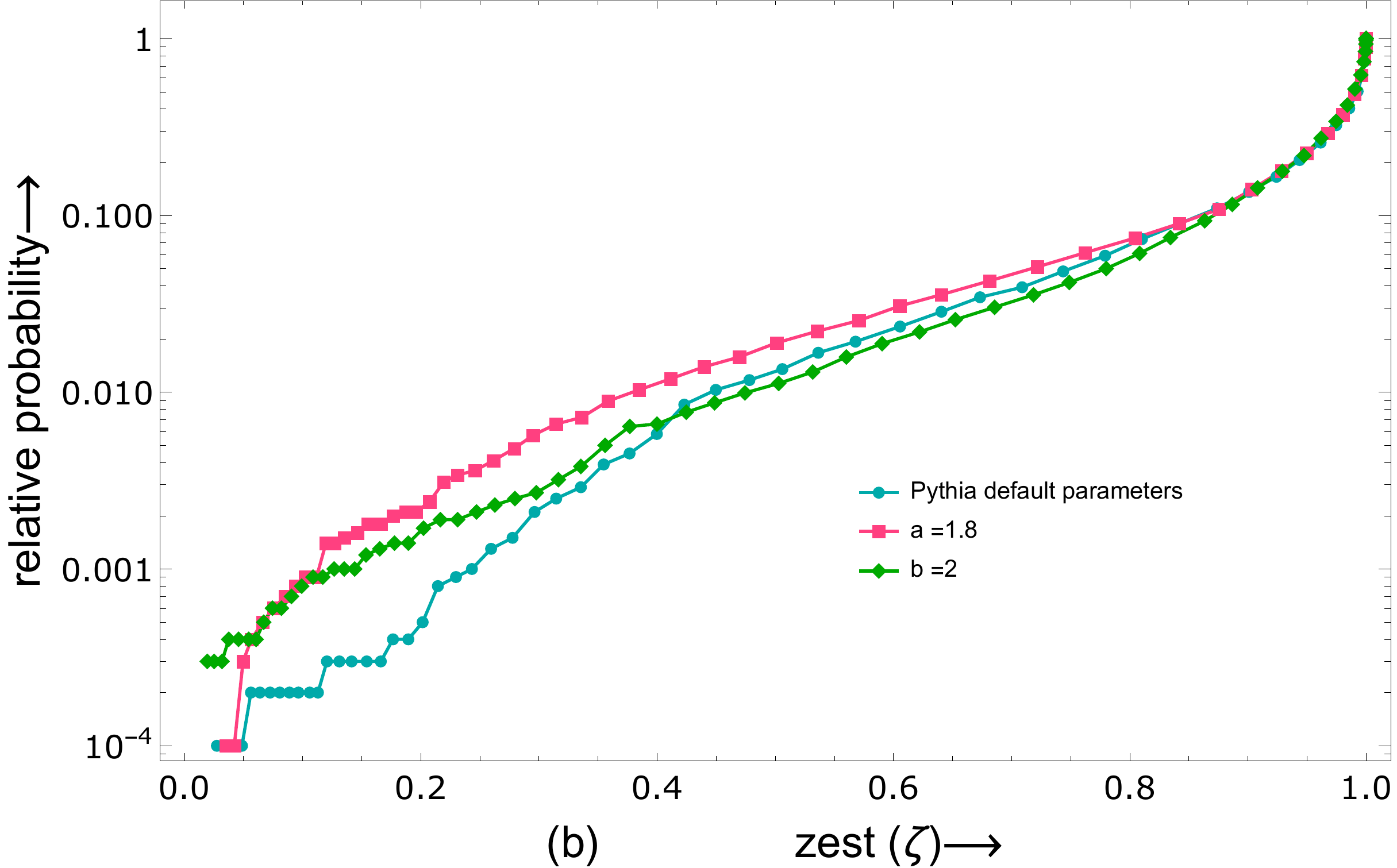}
 \caption{(a) Zest distribution curves for $Z$ boson-initiated jets with various choices of modelling the non-perturbative physics, and (b) ROC curves for $Z$ boson-initiated jets by varying the hadronization parameters in {\sc Pythia}.}
 \label{hadronization}
\end{figure*}
To study the hadronization model dependence of zest, we vary the non-perturbative physics modelling by: a) changing the hadronization parameters in {\sc Pythia}, and b) changing the hadronization model by using the {\sc Herwig} event generator to simulate the event. This will modify the spectrum of the final state primary hadrons, therefore allowing us to study the hadronization model dependence for zest.

In FIG.~\ref{hadronization} (a), we present the zest-distribution curves for a $Z$ boson-initiated jet with different clustering mechanisms for implementing the long distance physics. The {\sc Pythia} 8 event generator that we use for simulating the events, incorporates the Lund string model of hadronization \cite{Andersson:1983ia} to cluster final state partons into hadrons. This model is built upon a ``string" analogy, i.e. as the separation between the two partons increases, the potential energy stored in the string rises linearly. At some point, the potential energy stored is so large that the string breaks forming a $q\bar{q}$ pair leading to individual hadrons, with the `fragmentation function' given by,
\begin{equation}
  f(z) \propto z^{-1} (1-z)^{a} \exp(-b\, m_{\perp}^{2}/z)\, ,
  \label{lund}
\end{equation}
where $z$ is the energy fraction carried by the hadron and $m_{\perp}$ is its transverse mass while $a, b$ are some adjustable parameters.   
Changing $a$ and/or $b$ in the above equation, changes the way fragmentation happens, hence modifying the final state spectrum.\footnote{The default values for the parameters $a$ and $b$ used in {\sc Pythia} 8 are 0.3 and 0.8 respectively. Here
we will vary these values to their extremum limits by setting $a$ to 1.8 and $b$ to 2.} The zest-distribution shifts as shown in FIG.~\ref{hadronization}(a). Moreover, we observe a similar shift for the gluon zest distribution (background) and the performance curves change only very slightly, as shown in FIG.~\ref{hadronization}(b). For {\sc Herwig}, we cannot simulate single offshell gluons, thus we do not have corresponding curve for the gluon zest and ROC. However, from the heavy particle distribution for the {\sc Herwig} simulated event shown by the solid brown curve (with filled triangles) in FIG.~\ref{hadronization}(a), we see that the zest distribution lies well within the extreme bounds of hadronization parameters used in {\sc Pythia}, shown by the solid magenta (filled squares) and green (filled diamonds) curves in FIG.~\ref{hadronization}(a). 
Thus, from our analysis we conclude that although zest is not calculable perturbatively, but zest based discrimination is stable against different hadronization models.

Since zest is collinear unsafe, its discrimination ability may depend upon the resolution of the detector. We have also verified that while the zest curves shift if we coarse-grain the angular resolution between the particles for both the gluon-initiated and heavy-particle-initiated jets, they shift systematically such that the discrimination ability of the observable remains unaffected.

\section{Conclusion}
\label{Conclusion}

We have presented a new jet substructure observable, {\em zest}, and discussed its potential for discriminating standard model heavy particles forming jets from the QCD background of gluon-initiated jets. We have shown that the zest distribution of gluon-initiated jets is stable against the change in jet mass, change of global color flow of the partons and inclusion or exclusion of a few soft particles into/from the jet. These properties make it a suitable observable for vetoing the large gluon background at the colliders. Though zest is a non-linear and collinear unsafe observable, we have demonstrated that zest-based discrimination is largely insensitive to different hadronization models. We have shown that zest can be generalized through a real parameter, $p$, which can be optimized to further enhance the discrimination ability. A $p$ value between 0.3 to 0.5 is found to provide the optimal discrimination ability for all the heavy particle jets. We have also shown that for a top quark-initiated jet, $p$-zest provides a significant improvement over particle multiplicity in a jet. The discrimination provided by $p$-zest (with $p=0.3-0.5$) for the top quark-initiated jets approaches ML-based results, and we propose that looking at other IRC unsafe jet observables may help to uncover the physics hidden through such ML-based techniques.

\begin{acknowledgments}
We thank Tuhin Roy for discussions on some of the aspects of this paper and comments on the manuscript.
\end{acknowledgments}

\bibliographystyle{unsrt}
	\bibliography{Zest}

\end{document}